\documentclass[11pt]{article}
\usepackage{graphicx} 

\usepackage[bottom]{footmisc}

\usepackage[cm,plain]{fullpage}
\usepackage{amsmath}
\usepackage{epsf}
\usepackage{epsfig}

\usepackage{graphicx}
\usepackage{color}
\usepackage{latexsym}
\usepackage{amssymb}

\usepackage{a4wide}
\usepackage{xspace}
\usepackage{pslatex}
\usepackage{textcomp}

\usepackage{amsmath, amsthm, amssymb}

\usepackage[ ruled, lined, noresetcount]{algorithm2e}

\newtheorem{lemma}{Lemma}

\newtheorem{case_d}{Case Distinction}

\newtheorem{theorem}[lemma]{Theorem}
\newtheorem{proposition}[lemma]{Proposition}

\begin{document}

\newcommand{\NP}{\ensuremath{\mathcal{N}\mathcal{P}}\xspace}
\newcommand{\proofsketch}{\noindent\textbf{Proof sketch.}\enspace}
\newcommand{\bi}{\begin{itemize}}
\newcommand{\ei}{\end  {itemize}}
\newcommand{\bt}{\begin{tabbing}}
\newcommand{\et}{\end  {tabbing}}
\newcommand{\be}{\begin{enumerate}}
\newcommand{\ee}{\end  {enumerate}}
\newcommand{\bl}{\hspace*{2mm}}
\newcommand{\Bl}{\hspace*{5mm}}
\newcommand{\BL}{\hspace*{10mm}}
\newcommand{\sq}{\sqrt{2}}
\newcommand{\abs}[1]{{#1}}
\newcommand{\full}[1]{{}}

\title{Polygon Exploration with Time-Discrete Vision}
\author{
S\'andor P.\ Fekete\thanks{ Department of Computer Science,
Technische Universit{\"a}t Braunschweig, D-38106 Braunschweig,
Germany, Email: \{s.fekete,c.schmidt\}@tu-bs.de.
http://www.ibr.cs.tu-bs.de/alg/}
 \and Christiane Schmidt\footnotemark[1]
    \thanks{Supported by DFG Priority Program ``Algorithm Engineering'' (SPP 1307), project ``RoboRithmics'' (Fe 407/14-1). }
    }

\date{}
\maketitle


\maketitle

\markboth{}{}

\setcounter{footnote}{0}

\renewcommand{\abstractname}{Abstract:} 
\begin{abstract}

{\small With the advent of autonomous robots with two- and
three-dimensional scanning capabilities, classical visibility-based
exploration methods from computational geometry have gained in
practical importance. However, real-life laser scanning of useful
accuracy does not allow the robot to scan continuously while in
motion; instead, it has to stop each time it surveys its
environment.  This requirement was studied by Fekete, Klein and
N\"uchter for the subproblem of looking around a corner, but until
now has not been considered in an online setting for whole polygonal
regions.

We give the first algorithmic results for this important
optimization problem that combines stationary art gallery-type
aspects with watchman-type issues in an online scenario: We
demonstrate that even for orthoconvex polygons, a competitive
strategy can be achieved only for limited aspect ratio $A$ (the
ratio of the maximum and minimum edge length of the polygon), i.e.,
for a given lower bound on the size of an edge; we give a matching
upper bound by providing an $O(\log A)$-competitive strategy for
simple rectilinear polygons, using the assumption that each edge of
the polygon has to be fully visible from some scan point. }
\end{abstract}

{\bf Keywords:} Searching, scan cost, visibility problems, watchman
problems, online searching, competitive strategies, autonomous
mobile robots.

\section{Introduction}
{\bf Visibility Problems: Old and New.} The study of geometric
problems that are based on visibility is a well-established field
within computational geometry. The main motivation is guarding,
searching, or exploring a given region (known or unknown) by
stationary or mobile guards.

In recent years, the development of real-world autonomous robots has
progressed to the point where actual visibility-based guarding,
searching, and exploring become very serious practical challenges,
offering new perspectives for the application of algorithmic
solutions. However, some of the technical constraints that are
present in real life have been ignored in theory; taking them into
account gives rise to new algorithmic challenges, necessitating
further research on the theoretical side, and also triggering closer
interaction between theory and practice.

\begin{figure}
\vspace*{-3mm}
\begin{center}
  \includegraphics[width=40mm,height=40mm]{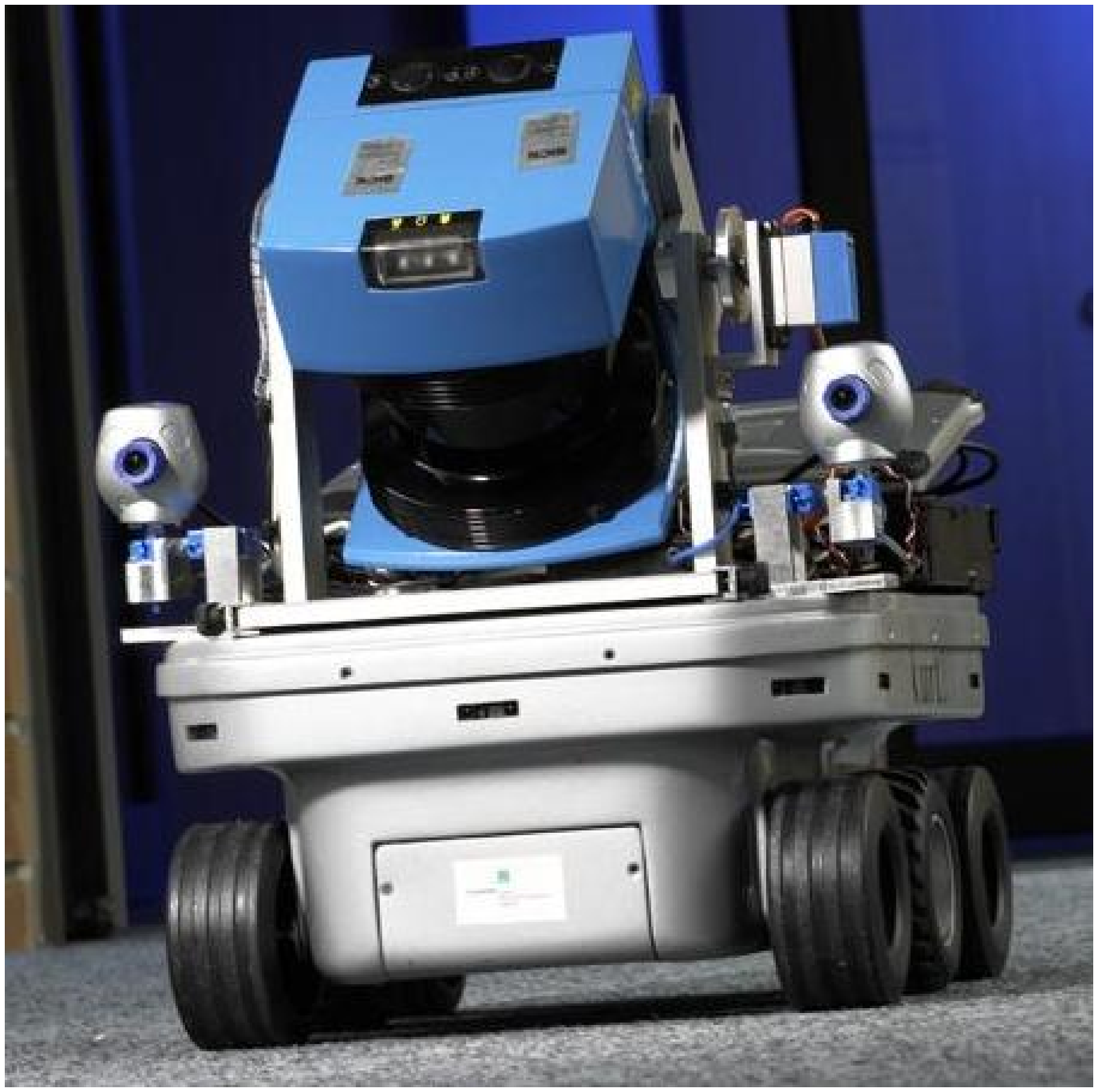}
  \includegraphics[width=40mm,height=40mm]{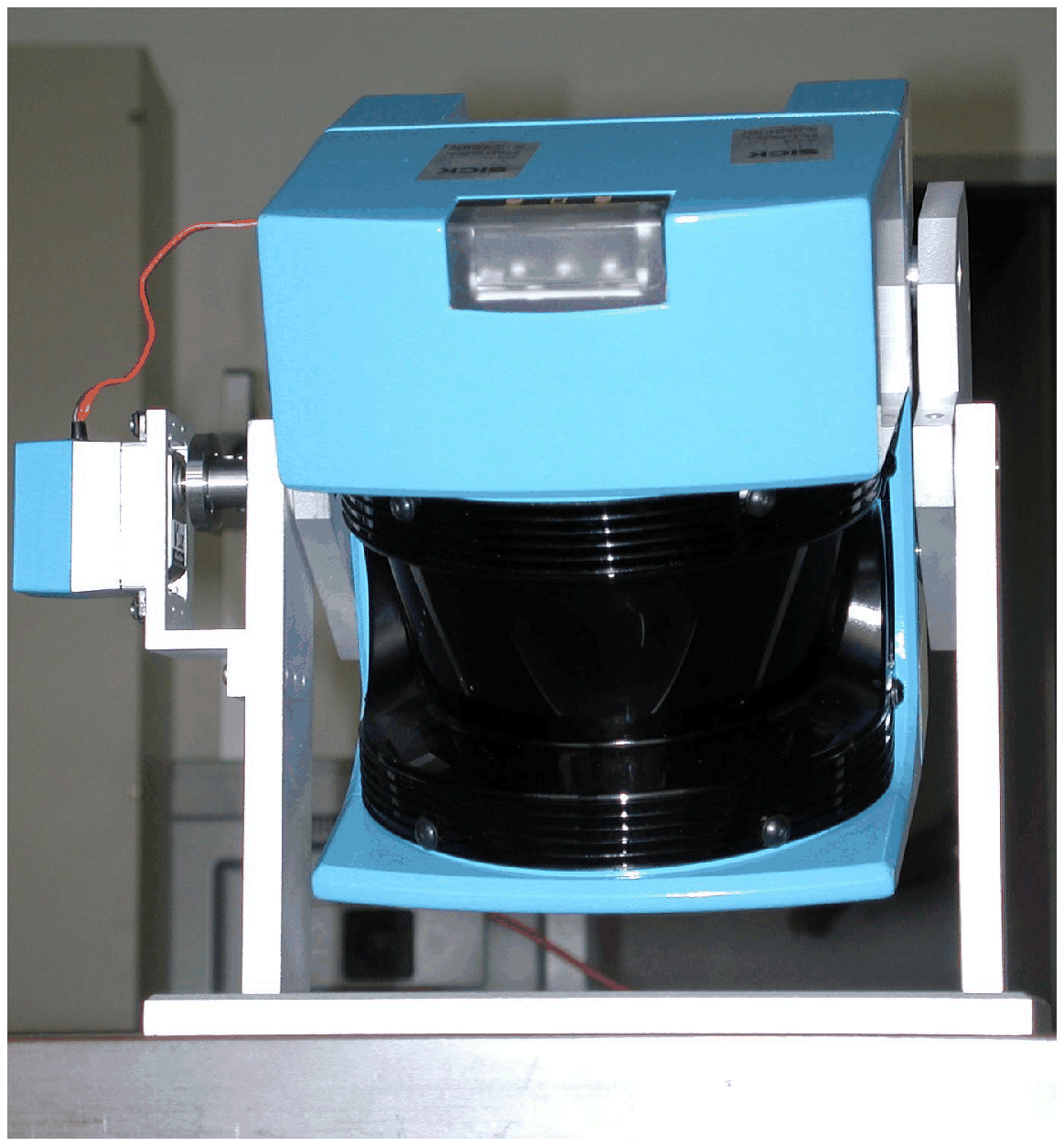}
\caption{Left: the autonomous mobile robot Kurt3D equipped with the
3D scanner. Right: the AIS 3D laser range finder. Its technical
basis is a SICK 2D laser range finder (LMS-200). (Both images used
with kind permission by Andreas N{\"u}chter, see \cite{RAAS2003}.)}
\label{dddscanner}
\end{center}
\vspace*{-3mm}
\end{figure}

One technical novelty that has lead to new possibilities and demands
is the development of high-resolution 3D laser scanners that are now
being used in robotics, see Figure~\ref{dddscanner} for an image and
\cite{RAAS2003} for technical details. By merging several 3D scans,
the robot Kurt3D builds a virtual 3D environment that allows it to
navigate, avoid obstacles, and detect objects~\cite{IAS2004}; this
makes visibility problems quite practical, as actually using good
trajectories is now possible and desirable. However, while human
mobile guards are generally assumed to have full vision at all
times, Kurt3D has to stop each time it scans the environment, taking
in the order of several seconds for doing so; the typical travel
time between scans is in the same order of magnitude, making it
necessary to balance the number of scans with the length of travel,
and requiring a combination of aspects of stationary art gallery
problems with the dynamic challenge of finding a short tour.

We give the first comprehensive study of the resulting {\em Online
Watchman Problem with Discrete Vision} (OWPDV) of exploring all of
an unknown region in the presence of a fixed cost for each scan. We
focus on the case of rectilinear polygons, which is particularly
relevant for practical applications, as it includes almost all
real-life buildings. We show that the problem is considerably more
malicious in the presence of holes than known for the classical
watchman problem; moreover, we demonstrate that even for extremely
simple classes of polygons, the competitive ratio depends on the
aspect ratio $A$ of the region; practically speaking, this
corresponds to the resolution of scans. Most remarkably, we are able
to develop an algorithm for the case of simple rectilinear polygons
that has competitive ratio $O(\log A)$, which is best possible; this
result uses the assumption that each edge of the polygon is fully
visibly from some scan point. This assumption has been used before
by Bottino and Laurentini~\cite{bl-nospabc-08}; in a practical
context, it can be justified by avoiding inaccuracies resulting from
putting together the individual scans. More importantly, it allows
it to sidestep a notorious open problem, which would otherwise come
up as a subproblem: finding a small set of stationary guards for a
polygonal environment. Details of this and other difficulties of
exploration with time-discrete vision are discussed in
Section~\ref{chap5}.

{\bf Classical Related Work.} Using a fixed set of positions for
guarding a known polygonal region is the classical {\em art gallery
problem}~\cite{Chv75,o-agta-87}. Schuchardt and
Hecker~\cite{sh-tnhag-95} showed that finding a minimum cardinality
set of guards is NP-hard, even for a simple rectilinear region; this
implies that the offline version of the minimum watchman problem
with discrete vision is also NP-hard, even in simple rectilinear
polygons.

Finding a short tour along which one mobile guard can see a given
region in its entirety is the {\em watchman problem}; see
Mitchell~\cite{joe.survey} for a survey. Chin and
Ntafos~\cite{chinta88} showed that such a watchman route can be
found in polynomial time in a simple rectilinear polygon, while
others \cite{thi-iacsw-93,thi-ciacs-99,cjn-fswrsp-99} found
polynomial-time algorithms for general simple polygons. Exploring
all of an unknown region is the {\em online watchman problem}. For a
simple polygon, Hoffmann et al.~\cite{hikk-pep-01} achieved a
constant competitive ratio of $c = 26.5$, while Albers et
al.~\cite{aks-eueo-99} showed that no constant competitive factor
exists for a region with holes and unbounded aspect ratio. For
simple rectilinear polygons, and distance traveled being measured
according to the Manhattan metric, the best known lower bound on the
competitive ratio is 5/4, as shown by Kleinberg~\cite{kleinberg}; if
distance traveled is measured according to the Manhattan metric,
Deng et al.~\cite{deng98how} gave an online algorithm for finding an
optimum watchman route (i.e. $c = 1$) in case a starting point on
the boundary is given (otherwise $c = 2$; the best known upper bound
in this case is $c = 3/2$ \cite{hnp-cerp-06}). Note that our
approach for the problem with discrete vision is partly based on
this GREEDY-ONLINE algorithm, but needs considerable additional
work.

Another online scenario that has been studied is the question of how
to look around a corner: Given a starting position, and a known
distance to a corner, how should one move in order to see a hidden
object (or the other part of the wall) as quickly as possible? This
problem was solved by Icking et al.~\cite{ikm-hlac-93,ikm-ocsla-94},
who show that an optimal strategy has competitive factor of
1.2121\ldots.

{\bf Searching with Discrete Penalties.} In the presence of a cost
for each discrete scan, any optimal tour consists of a polygonal
path, with the total cost being a linear combination of the path
length and the number of vertices in the path. A somewhat related
problem is searching for an object on a line in the presence of turn
cost~\cite{dfg-ostc-06}, which turns out to be a generalization of
the classical linear search problem.

Somewhat surprisingly, scan cost (however small it may be) causes a
crucial difference to the well-studied case without scan cost, even
in the limit of infinitesimally small scan times: Fekete et
al.~\cite{fkn-osar-06} have established an asymptotically optimal
competitive ratio of 2 for the problem of looking around a corner
with scan cost, as opposed to the optimal ratio of
1.2121\ldots without scan cost, cited above. 

{\bf Other Related Work.} Visibility-based navigation of robots
involves a variety of different aspects. 
For example, Carlsson and Nilsson \cite{carlsson99computing} give an
efficient algorithm to solve the problem of placing stationary
guards along a given watchman route, the so-called vision point
problem, in streets.
Ghosh et al.~\cite{gb-euped-07, gbbs-oadveupe-08} study unknown
exploration with discrete vision, but they focus on the worst-case
number of necessary scan points (which is shown to be $r+1$ for a
polygon with $r$ reflex vertices), their algorithm results in a (not
constant) competitive ratio of $(r+1)/2$, and on scanning along a
given tour, without deriving a competitive strategy. For the case of
a limited range of visibility Ghosh et al.~give an algorithm where
the competitive ratio in a Polygon $P$ (with $h$ holes) can be
limited by $\lfloor\frac{8\pi}{3} + \frac{\pi R \times
Perimeter(P)}{Area(P)} + \frac{(r+h+1)\times \pi
R^2}{Area(P)}\rfloor$.

{\bf Our Results.} We give the first comprehensive algorithmic study
of visibility-based online exploration in the presence of scan cost,
i.e., discrete vision, by considering an unknown polygonal
environment. This is interesting and novel not only in theory, it is
also an important step in making algorithmic methods from
computational geometry more useful in practice, extending the
demonstration from the video~\cite{fkn-sarv-04}.

After demonstrating that the presence of discrete vision adds a
number of serious difficulties to polygon exploration by an
autonomous robot, we present the following mathematical results:

\vspace*{-.3cm}
\begin{itemize}
\item We demonstrate that
a competitive strategy is possible only if maximum and minimum edge
length in the polygon are bounded, i.e., for limited resolution of
the scanning device. More precisely, we give an $\Omega(\log A)$
lower bound on the competitive ratio that depends on the aspect
ratio $A$ of the region that is to be searched; the aspect ratio $A$
is the ratio of maximum and minimum edge length. If the input size
of coordinates is not taken into account, we get an $\Omega(n)$
lower bound on the competitive factor. This bound is valid even for
the special case of orthoconvex polygons, which is extremely simple
for continuous vision.
\item For the natural special case of simple rectilinear
polygons (which includes almost all real-life buildings), we provide
a matching competitive strategy with performance $O(\log A)$.
\end{itemize}

The rest of this paper is organized as follows. Section
\ref{sec:prel} presents some basic definitions and the basic ideas
of a strategy for continuous vision. A number of additional
difficulties for discrete vision are discussed in
Section~\ref{chap5}. In Section~\ref{sec:aspect} we demonstrate that
even very simple classes of polygons (orthoconvex polygons with
aspect ratio $A$) require $O(\log A)$ scans. On the positive side,
Section~\ref{sec:mathfound} lays the mathematical foundations for
the main result of this paper, which is presented in
Section~\ref{sec:strategy}: an $O(\log A)$-competitive strategy for
the watchman problem with scan costs in simple rectilinear polygons.
The final Section~7 provides some directions for future research.

\section{Preliminaries}\label{sec:prel}
\subsection{Definitions}\label{subsec:def}
Let $P$ be a simple rectilinear polygon, a polygon without holes and
internal angles of either 90 or 270 degrees at all vertices. Two
points $p$ and $q$ in $P$ are {\it visible} to each other in case
the line segment connecting $p$ and $q$ lies inside of $P$.
Moreover, a polygon is said to be \textit{monotone} (with respect to
a given line L), if any intersection with a line that is orthogonal
to L is an interval, i.e., the intersection is either a line segment
or a single point or the empty set. A vertex of $P$ is a {\it reflex
vertex} if the internal angle is larger than 180 degree. Hence, for
a rectilinear polygon the vertices with an internal angle of 270
degree are reflex.

Considering an edge $e$ of the polygon, the {\it weak visibility
polygon} of $e$ consists of all points that see at least one point
of $e$. The points in the {\it integer visibility polygon} see all
points of $e$. As we demand that each edge is fully visible from one
scan point we need at least one scan point in the integer visibility
polygon of each side.

In the following we will measure the length of the tour according to
the {\it $L_1$ metric}.(That is, the distance between two points $p$
and $q$ is given by: $ ||p-q||_{L_1} = |p_x - q_x| + |p_y - q_y|$;
here $p_x$ and $p_y$ denote the x- and y-coordinate of a point $p$.)

\subsection{Extensions}\label{subsec:exts}
The use of extensions is a central idea of polygon exploration
(see~\cite{deng98how}). Each extension is induced by one or two
sides of the polygon $P$. More precisely, at each reflex vertex we
extend each side $S$ of $P$ inside the polygon until this line hits
the boundary of $P$. We obtain a line segment excluding $S$; this is
called an \textit{extension} of $S$.
For structuring the set of all extensions, the notion of {\em
domination} turns out to be useful, giving rise to different types
of extensions as follows. From a starting point of the robot, any
extension $E$ of a side $S$ divides the polygon into two
sub-polygons: a polygon including the starting point, and the other
subpolygon FP[$E$] (the \textit{foreign polygon defined by E}).
There exist sides $S$$\in$FP[$E$] which become visible for the robot
only if $E$ is \textit{visited}, i.e., if the robot either crosses
or touches the extension. As we want to explore the entire polygon,
$S$ must be visible at some point of the tour; therefore, visiting
$E$ is necessary for exploring $P$, which is why we call such an
extension \textit{necessary}. Moreover, it is possible that for two
necessary extensions $E_1$ and $E_2$ the robot cannot reach $E_1$
without crossing $E_2$, as FP[$E_2$] contains all of $E_1$. As we
will visit (even cross) $E_2$ when we visit $E_1$, we may
concentrate on $E_1$. In this case $E_1$ dominates $E_2$. A
nondominated extension is called an \textit{essential extension}.

\subsection{GREEDY-ONLINE}
\label{subsec:greedy} The GREEDY-ONLINE algorithm of Deng et al.\
\cite{deng98how} deals with the online watchman problem in simple
rectilinear polygons for a robot with continuous vision. The basic
idea of this algorithm is to identify the clockwise bound of the
currently visible boundary; subsequently they consider a necessary
extension that is defined either by the corner incident with this
bound, or by a sight-blocking corner. This is based on a proposition
of Chin and Ntafos \cite{chinta88}: there always exists a
noncrossing shortest path, i.e., a path that visits the critical
extensions in the same circular order as the edges on the boundary
that induce them. It is vital to establish a similar property for
the case of discrete scans.

Chin and Ntafos \cite{chinta88} started with optimum watchman routes
in monotone rectilinear polygons, then extended this to rectilinear
simple polygons. Without loss of generality, Chin and Ntafos
presumed the edges to be either vertical or horizontal, and
monotonicity referring to the $y$-axis. They called an edge on the
boundary a \textit{top edge}, if the interior of the polygon is
located below it. Analogously, a \textit{bottom edge} is an edge
below the interior of the polygon. The highest bottom edge is named
$T$, the lowest top edge $B$. The part of the polygon that lies
above $T$ is called $P_t$. Considering the kernel of $P_t$, the {\it
top kernel}, i.e., the part of it that can see every point of $P_t$,
Chin and Ntafos named its bottom boundary $K_t$. Analogously, $P_b$,
$K_b$ and the {\it bottom kernel} are defined as the part of the
polygon that is located below $B$, the top boundary of $P_b$'s
kernel and the kernel of $P_b$.

\section{Difficulties of Time-Discrete Vision}\label{chap5}
The main result of this paper is to develop an exploration strategy
for simple rectilinear polygons. Our approach will largely be based
on the strategy GREEDY-ONLINE by Deng, Kameda, and Papadimitriou
\cite{deng98how}, which is optimal (i.e., 1-competitive) in $L_1$
for continuous vision and a given starting point on the boundary
(Section~\ref{subsec:greedy}). That algorithm itself is based on
properties first established by Chin and Ntafos~\cite{chinta88},
focusing on critical extensions; we will describe further details in
Section~\ref{sec:mathfound} and \ref{sec:strategy}.

A basic difficulty we face when developing a good online strategy is
the reference to an optimum: A robot with continuous vision simply
has to keep an eye on the tour length. The combination with scan
costs makes it much harder to have a benchmark for comparison, as we
have to balance both tour length and number of scans. This becomes
particularly challenging when facing a variety of geometric issues,
illustrated in the following.

First and foremost, finding an optimal tour requires determining a
small set of stationary guard positions that completely covers the
complete interior of a polygon. Even in an offline scenario, this is
notoriously difficult; at this point, the best known offline
approximation algorithms yield $O(\log n)$-approximations, e.g.,
with run time $O(n^4)$ for simple polygons by an improved version of
\cite{g-aaagp-87}.

\begin{figure}[h]
\centering
\epsfig{file= 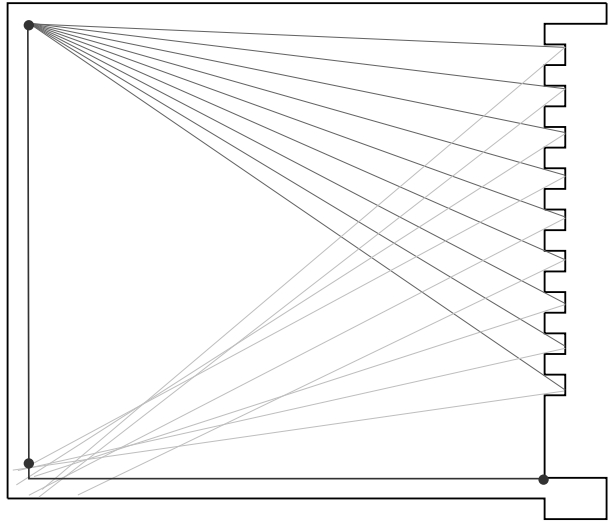, width=8cm } \\
\caption{\label{fig:ag} The niches on the right side of the polygon
can be guarded using only two scan points.}
\end{figure}

The main difficulty is illustrated in Figure~\ref{fig:ag}: The
niches on the right can be covered using only two scan points;
neither of them covers a whole niche, and neither is chosen from an
obvious set of discrete candidates for guard positions. This issue
can be side-stepped by our assumption that each edge needs to be
fully visible from some scan point.

\begin{figure}[h]
\hspace{0.2cm} (a) \hspace{0.5cm}
\begin{minipage}[h]{3.5cm}
\begin{center}
   \epsfxsize=2.8 cm
   \raisebox{1.2ex}{$s$}\hspace{0.05cm}\epsfbox{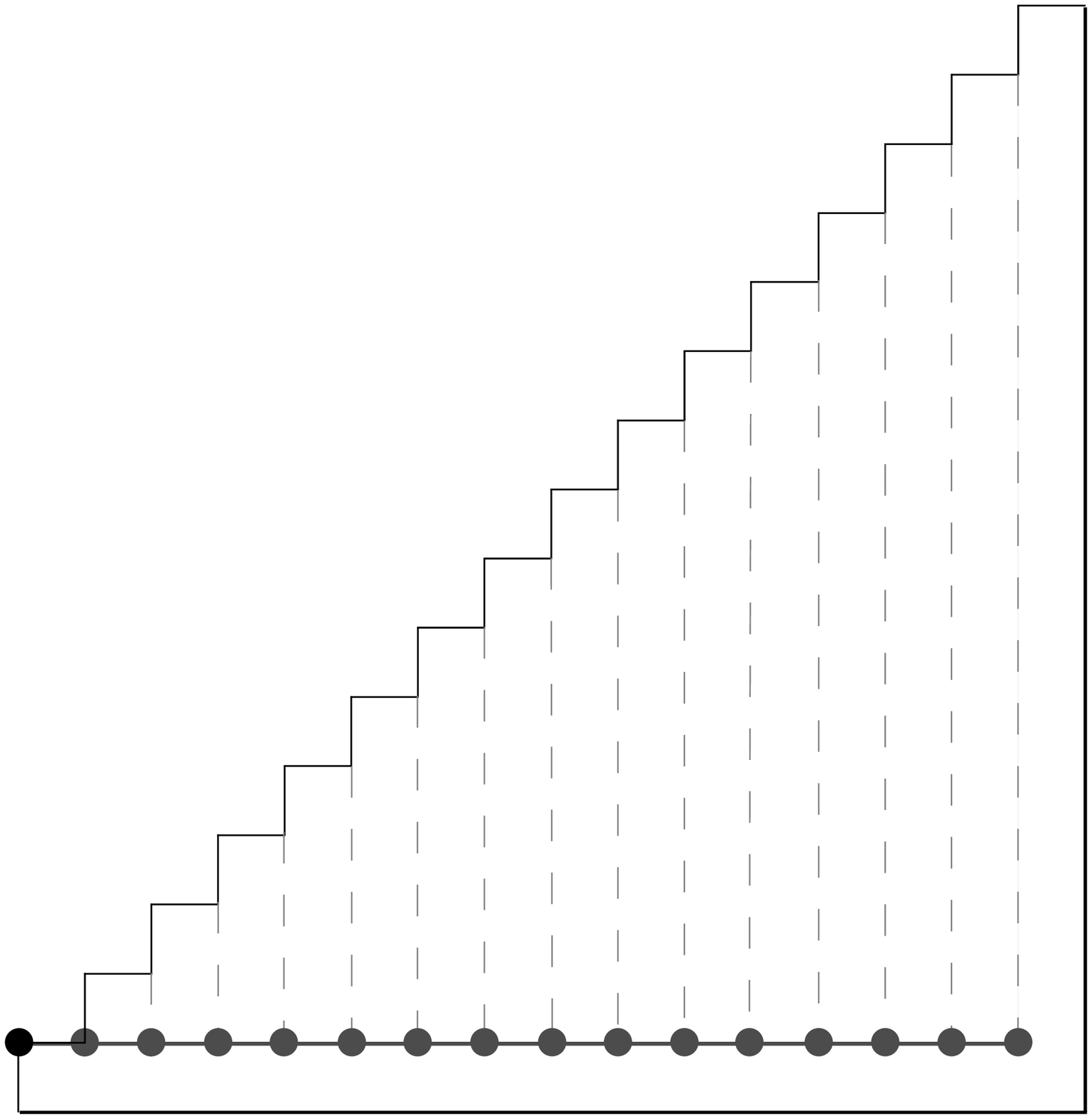}
\end{center}
\end{minipage}
\begin{minipage}{3.5cm}
\begin{center}
   \epsfxsize=2.8 cm
   \raisebox{1.2ex}{$s$}\hspace{0.05cm}\epsfbox{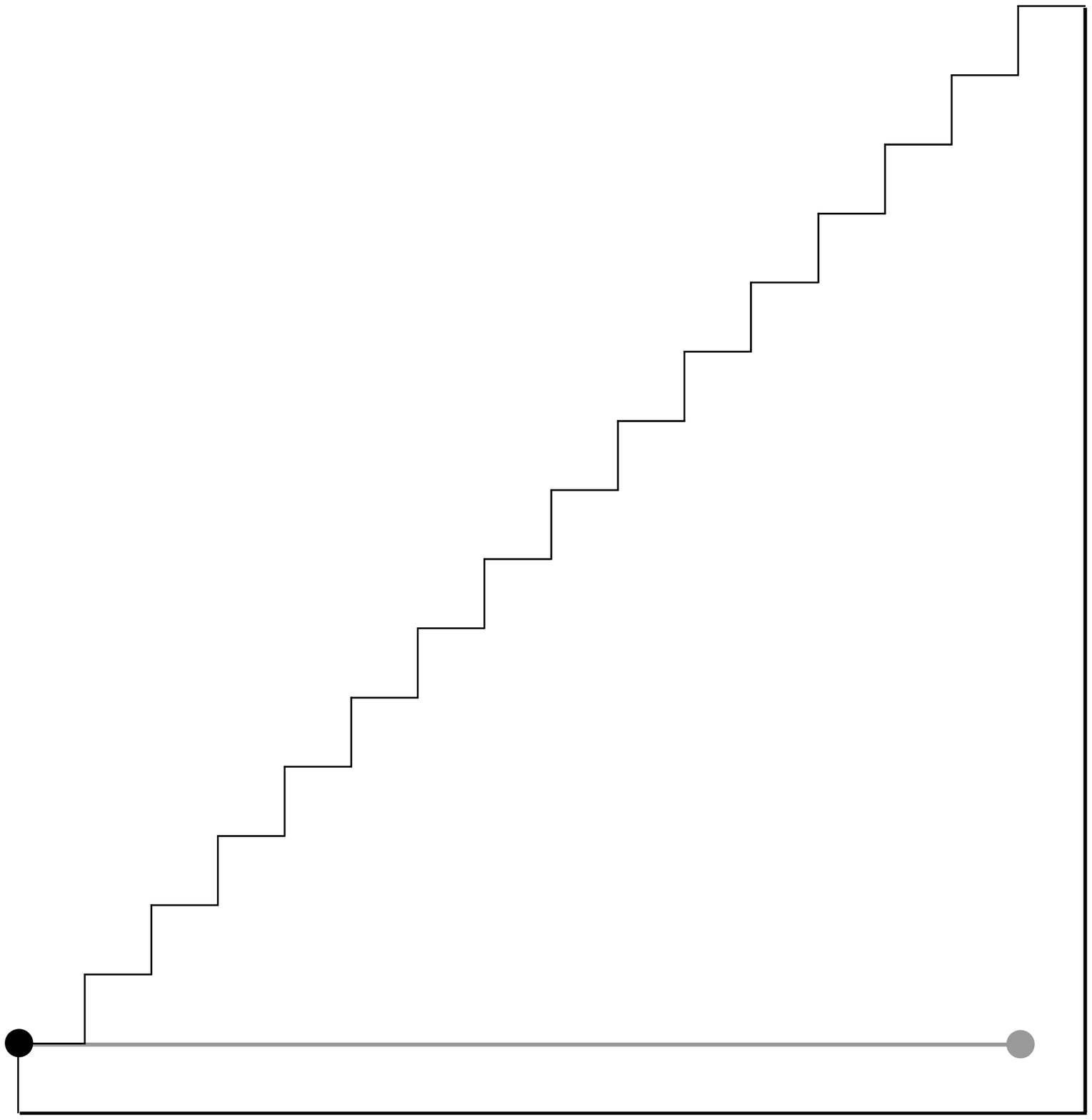}
\end{center}
\end{minipage}
\hspace{1.9cm}(b) \hspace{0.1cm}
\begin{minipage}{2.0cm}
\begin{center}
   \epsfxsize=1.2 cm
   \raisebox{0.8ex}{$s$}\hspace{0.05cm}\epsfbox{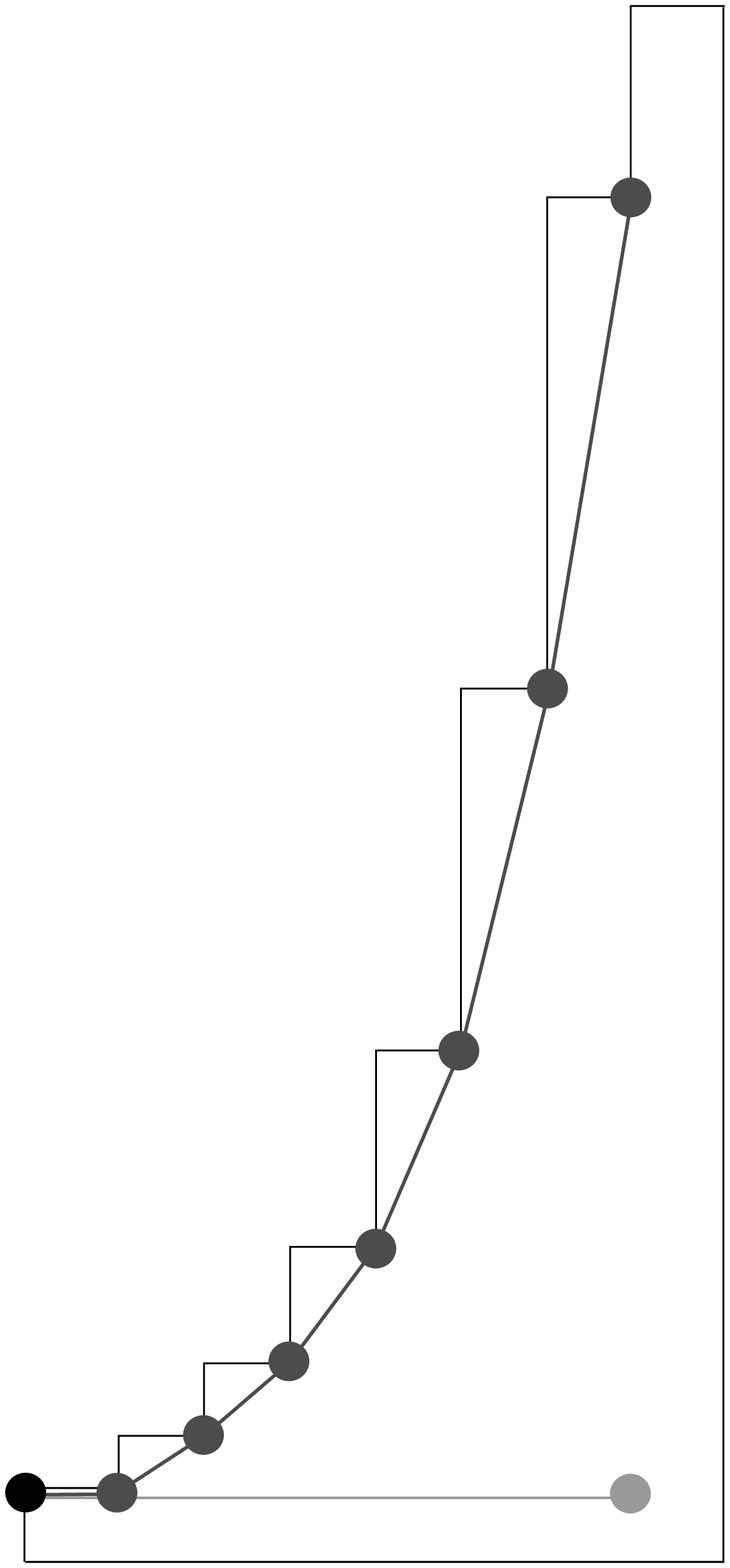}
\end{center}
\end{minipage}\bigskip\\
\hspace*{0.2cm} (c)
\begin{minipage}{8.6cm}
\begin{center}
   \epsfxsize=8.1 cm
   \raisebox{1.8ex}{$s$}\hspace{-0.15cm}\epsfbox{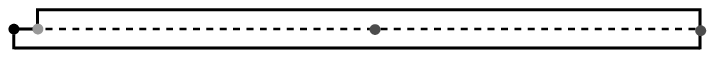}
\end{center}
\end{minipage}
\hspace{1cm}(d)
\begin{minipage}{2.7cm}
\begin{center}
   \epsfxsize=1.8 cm
   \epsfbox{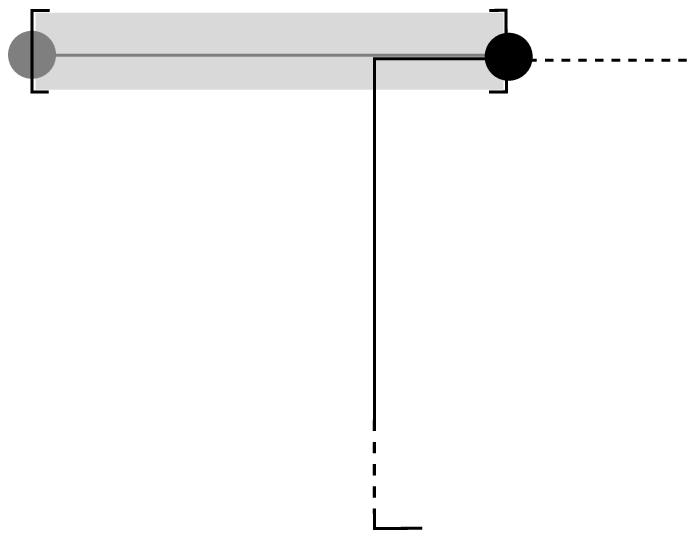}\hspace{-0.35cm}\raisebox{9ex}{$s$}
\end{center}
\end{minipage}

 \caption{\label{fig:stepchoice} A number of problems faced by a strategy; the starting point is given by $s$, 
 necessary extensions are drawn dashed.
(a) If too many necessary extensions are taken into account, the
strategy (left) may end up using a large number of scans, while the
optimum takes only one scan (right). (b) When going for next visible
corners (dark-gray), the (gray) optimum may not have to leave the
base line, causing an arbitrarily bad performance. (c) Going too far
beyond a corner may also end up being bad: The optimum needs only a
step to the corner and a single scan to see the entire polygon. (d)
If the gray line represents the next planed step, the optimum has
the opportunity to turn off at some point in the light gray
interval. }
\end{figure}

Even then, a number of problems are faced by an online strategy:

 \textbf{(a) Scanning too often.} As
opposed to the situation with continuous vision, our strategy needs
to avoid too many scans. When simply focusing on edge extensions, we
may run into the problem shown in Figure~\ref{fig:stepchoice}(a);
this also highlights the problem of not knowing {\em critical}
extensions before they have been visited. That is, they may not be
distinguished from necessary extensions, but scanning on each such
one may cause an arbitrarily high competitive ratio.

\textbf{(b) Where to go next.} The robot faces another dilemma when
choosing the next scan point: should it walk to the next corner (or
the next chosen reference point) itself---or to its perpendicular?
Going for the next corner may cause a serious detour,
see~Figure~\ref{fig:stepchoice}(b).

\textbf{(c) Missing a scan.} As seen in
Figure~\ref{fig:stepchoice}(a), it is not a good idea to stop at
each corner; on the other hand, when facing a corner in a certain
distance and an unknown area behind it, using a predefined point in
the unknown interval, e.g., the center or the end, does not allow a
bounded competitive ratio, see~Figure~\ref{fig:stepchoice}(c).

\textbf{(d) Missing a turn.} Searching for the right distance to
place a scan may cause the robot to run beyond an extension, while
the optimum may have the opportunity to turn off earlier: e.g.,
consider the shaded interval in Figure~\ref{fig:stepchoice} (d).
This makes it necessary to consider adjustments and also holds for
other situations in which an early turn of the optimum may be
possible.

\begin{figure}[h]
\centering \epsfig{file=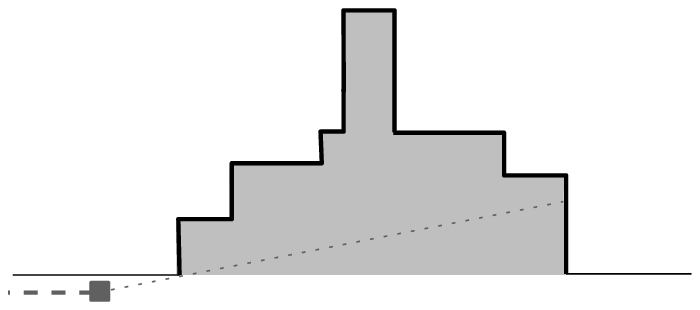,width=.30\textwidth}\hspace{1cm}
\epsfig{file=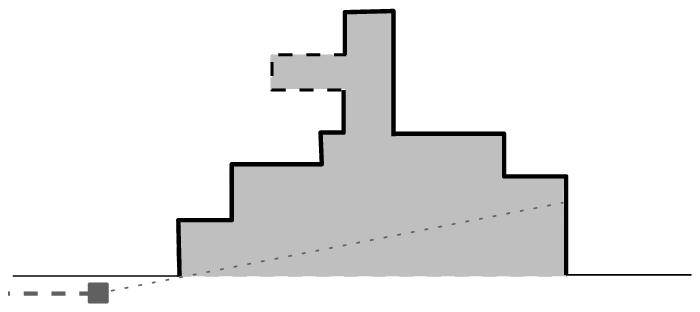,width=.30\textwidth} \caption{ \label{nvr} Left:
If the dark gray point represent the scan position, a robot with
discrete vision cannot see the entire bold sides, resulting in a
non-visible region (NVR), shown in gray; an NVR is dealt with by
performing a binary search. Scans located in the dotted interval
allow for the perception of all sides of the NVR. Right: Within a
non-visible region, there may be parts that even a robot with
continuous vision may not see completely (dashed), requiring the
robot to enter the NVR; such a situation is dealt with by
introducing turn adjustments, when the need arises.}
\end{figure}

\textbf{Non-visible regions.} Even situations that are trivial for a
robot with continuous vision may lead to serious difficulties in the
case of discrete vision, as illustrated in Figure~\ref{nvr}(left):
Without entering the gray area, a watchman with continuous vision is
able to see the bold sides completely. A robot with discrete vision
is able to see these bold parts of the boundary if only it chooses a
scan point under the northernmost part of the boundary. Such an area
for which not (yet) all sides which would be completely visible with
continuous vision (the bold sides) are visible for a robot with
discrete vision, is called a \textit{non-visible region} (NVR).

All in all, serious adjustments have to be made to establish
mathematical structure for exploration with discrete vision; this
work is presented in Section~\ref{sec:mathfound}. Enhanced by
several important additional insights and tools (sketched in
Sections~\ref{subsec:binary} and \ref{subsec:turn}), we get our
strategy SCANSEARCH, which is presented in
Sections~\ref{subsec:decisions}, \ref{subsec:wtg}, \ref{subsec:dtam}
, \ref{subsec:htm} and~\ref{subsec:strategy}. The resulting
competitive ratios are discussed in Section~\ref{subsec:scansearch}.

\section{Why the Aspect Ratio Matters}\label{sec:aspect}
Before developing the details of our $O(\log A)$-competitive
strategy for a simple orthogonal polygon with aspect ratio $A$, we
illustrate that this is best possible, even for orthoconvex polygons
that contain a single niche, given by a staircase to its left and to
its right as shown in Figure~\ref{fig:niche}.

\begin{theorem}
Let $P$ be a polygon with $n$ edges and aspect ratio $A$. Then no
deterministic strategy can achieve a competitive ratio better than
$\Omega(\log{A})$, even if $P$ is known in advance to be an
orthoconvex polygon.
\end{theorem}

\begin{proof}
In the beginning, the robot with discrete vision stands at some
point with distance $\delta$ to the base line of the niche and small
distance $d$ to the perpendicular of one of the corners $a_0, b_0$
(w.l.o.g.\ $a_0$). In an optimal solution, a single scan suffices to
see the entire polygon, provided it is taken within the strip shaded
in gray. However, the robot does not know the location of this
strip, as it depends on reflex vertices of the polygon that are not
visible yet. More precisely, let [$a_0, b_0$] be the initial
interval.  We divide each interval [$a_i, b_i$] into three intervals
of equal length; only one of the outermost intervals is open, the
other two coincide with the boundary. This defines the new interval
[$a_{i+1}, b_{i+1}$]; see the middle of Figure \ref{fig:niche}. Let
$x_i$ be the position of the robot in the corresponding $i$th
interval.

In order to show a lower bound on the competitive strategy we have
to construct a scene such that, no matter how the online strategy
chooses the next location a certain number of steps cannot be
avoided. The scene is constructed by the ``adversary''---responding
to the behavior of the strategy. Here, choosing the next scan point
closer to one boundary, the adversary leaves the other outermost
interval open. When choosing the next location in the center of the
interval the adversary makes sure with the length of the vertical
sides that no information on the layout of the next step is gained.

\begin{figure}[h]
\centering
\epsfig{file= 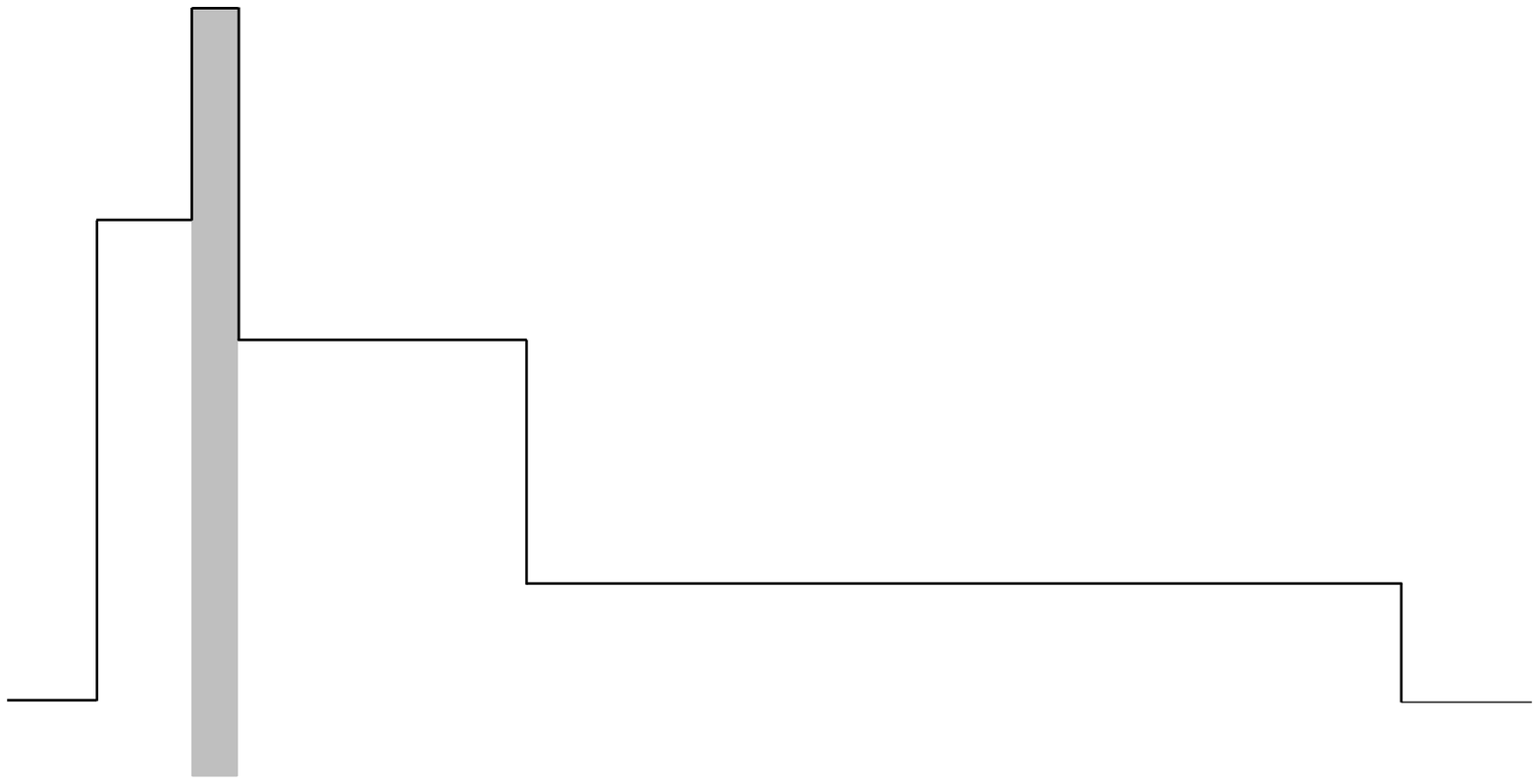, width=4cm } \hspace{1cm}\epsfig{file= 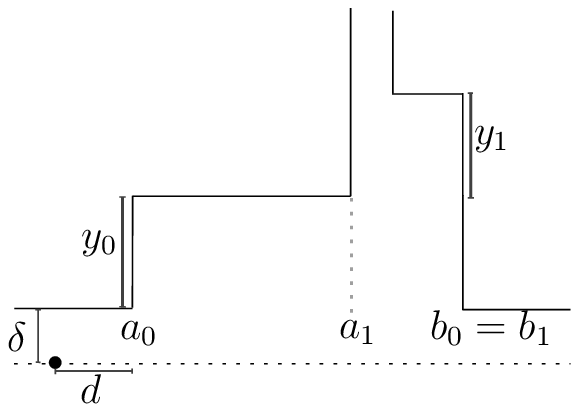, width = 4cm }\hspace{1cm} \epsfig{file=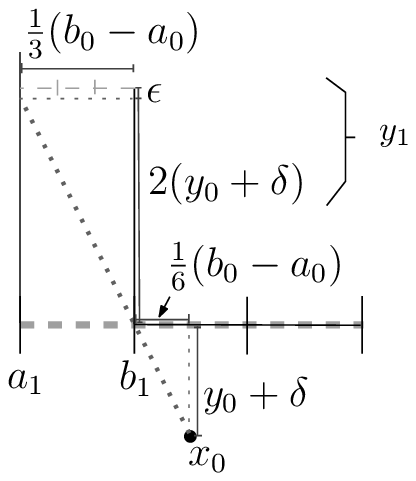,width=3cm} \\
\caption{\label{fig:niche}Left: One scan in the gray strip suffices
to see the entire niche; middle: definition of $\delta, d, a_i, b_i,
y_i$; right: computing $y_1$ to bound the coordinate values.}
\end{figure}

Choosing $y_0 = \delta/d (b_0-a_0) + \varepsilon$, $y_1 =
2(y_0+\delta)+\varepsilon, \ldots$ the robot cannot see the entire
side when located at $x_i$. This results in an exponential lower
bound for the aspect ratio: the smallest side length in the $i$th
step of our construction is
$(b_i-a_i)=\left(\frac{1}{3}\right)^{i}(b_0-a_0)$. For the maximum
side length we have $2\delta/d3^{i}(b_0-a_0)+T$, $T<f3^{i}(b_0-a_0)$
with some constant $f$. Consequently: $i = 1/2 \cdot \frac{\log A -
\log(2\delta/d+f)}{\log 3}$.

Thus, for any given
aspect ratio $A$, a total number of $\Omega(\log A)$ scans cannot be
avoided to guarantee full exploration of the niche. Note that the
total number of scans can also be described as $\Omega(n)$; however,
this lower bound is not purely combinatorial, as it depends on the
coding of the input size.
\end{proof}

\section{Mathematical Foundations}\label{sec:mathfound}
In the following, we will deal with a limited aspect ratio by
assuming a minimum edge length of $a$; for simplicity, we assume
that the cost of a scan is equal to the time the robots needs for
traveling a distance of 1. Hence, we concentrate in the following on
$a < 1$. Moreover, we still concentrate on orthoconvex polygons.

The correctness of the GREEDY-ONLINE algorithm is based on the
propositions of Chin and Ntafos~\cite{chinta88}, Section
\ref{subsec:greedy}.
 When considering discrete vision, even the simplest proposition
on monotone rectilinear polygons breaks down: Finding an optimum
watchman route is {\em not} necessarily equivalent to finding a
shortest path connecting the top and bottom kernels, as we need to
take into account that some scans have to be taken along the way.
The cost  $t(T)$ of our tour $T$ is a linear combination of the tour
length and the number of scan points (in the set {\cal S} of scan
points) used along this route: $t(T) = c\cdot |{\cal S}| + L(T)$.
Thus, {\it shortest} refers to a tour with lowest cost. In the
following, we will develop several modifications for discrete vision
of increasing difficulty that lay the foundation for our algorithm
SCANSEARCH.

In the following, a \textit{visibility path} is a path with scans,
along which the same area is visible for a robot with discrete
vision, as it would be for a human guard with continuous vision. We
will proceed by a series of modifications to the results by Chin and
Ntafos; modifications are highlighted, and the numbering in
parentheses with asterisks refers to that in \cite{chinta88}.

\begin{lemma}[Lemma 1*]\label{Lemma1neu}
Finding an optimum watchman route \textbf{of a robot with discrete
vision} in a monotone rectilinear polygon is equivalent to finding a
shortest \textbf{visibility} path that connects the top and bottom
kernels. In case of a polygon that is monotone two both axes the
intersection of the kernels is to be considered, cp.~Figure
\ref{xymonotone}.
\end{lemma}

\begin{proof} We distinguish two cases. If the given polygon is star-shaped,
the top and the bottom kernel coincide, and any point in the kernel
is an optimum watchman route for a robot.

If the given polygon is not star-shaped, and hence the top and
bottom kernel do not coincide, a shortest visibility path between
$K_t$ and $K_b$ is an optimum watchman route:

\begin{itemize}
  \item[-] First no optimum watchman route of a robot with discrete vision
extends above $T$ (the highest bottom edge) or below $B$ (the lowest
top edge); in that case, we could find a shorter route as follows.

If the path to the point above $T$ or below $B$ and the one that
leaves that point are the same beyond $T$ or $B$, we move the last
scan point to $T$ or $B$ and cut off the end of the route.

Or, if there is an angle greater than $0$ between the in- and
outgoing path, we construct a shorter route by moving the last scan
point to $T$ or $B$ (to the point with the shortest distance) and
connect the new point with the next scan points.
  \item[-] Let $S$ be a shortest visibility path from $K_t$
  (the bottom boundary of the kernel of $P_t$,
the portion of the polygon that lies above $T$,) to $K_b$. Every
point in the polygon is visible from some point along $S$: $P_t$ and
$P_b$ are visible from the endpoints of $S$, which lie on $K_t$ or
$K_B$; a point elsewhere in the polygon must be visible because it
is visible from the corresponding path of a robot with continuous
vision (because of the monotonicity) and the definition of a
visibility path.

Then an optimum watchman route of a robot with discrete vision is
formed by following this shortest visibility path and walking
backwards (without a scan if possible, i.e., if no shift in
direction is needed).
\end{itemize}

\begin{figure}[h]
\centering
\epsfig{file= 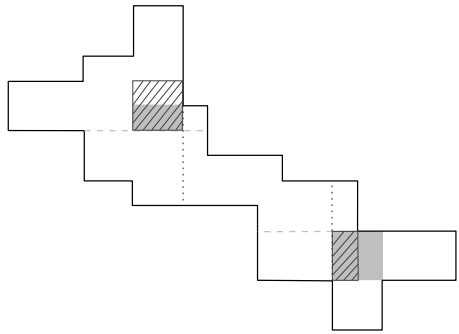, width=6cm } \\
\caption{\label{xymonotone}A polygon that is monotone wrt the x- and
the y-axis. The top and bottom kernels are shown in light gray and
in shaded dark gray.}
\end{figure}

\end{proof}

Just like Chin and Ntafos \cite{chinta88}, we now focus on
rectilinear simple polygons and adopt their procedure, i.e., we
first partition the polygon into uniformly monotone rectilinear
polygons and then identify for each of the resulting polygons $R_i$
the bottom edges of top kernels ($T_i$) and the top edges of bottom
kernels ($B_i$). First, we identify the essential horizontal edges,
and, after applying the method to the polygon after a 90 degrees
rotation, the essential vertical edges.

Like in the case of monotone rectilinear polygons, the portions of
the polygons that lie outside of the essentials edges will not be
visited by any optimum watchman route of the considered robot and
are discarded.

\begin{lemma}[Lemma 2*]\label{Lemma2neu}
If $P$ is the original rectilinear simple polygon and $P'$ is the
new polygon obtained by removing the "non-essential" portions of the
polygon, then no optimum watchman route \textbf{of a robot with
discrete vision} will visit any point in $P\setminus P'$.
\end{lemma}

\begin{proof} If the claim were not true, i.e., there would be an optimum watchman
route of a robot with discrete vision visiting a point in
$P\setminus P'$, this route would cross at least one essential edge.
Any point in the section of that edge that is enclosed by the route
can see the portion of the polygon that is in $P\setminus P'$ so we
can make the route shorter, as we would need at least one scan in
$P\setminus P'$ for the former route as well. Thus, we have a
contradiction to the proposition that we have an optimum watchman
route. \end{proof}

This allows us to reformulate Lemma 3 of Chin and Ntafos:

\begin{lemma}[Lemma 3*] \label{Lemma3neu}
There is an optimum watchman route \textbf{of a robot with discrete
vision} in $P$ that visits the essential edges in the order in which
they appear on the boundary of $P'$.
\end{lemma}

\begin{proof}
If an optimum watchman route of a robot with discrete vision does
not visit the essential edges in this order, the route will
intersect itself.

Then we can restructure this route by deleting this intersection and
get a shorter route in which the pre-specified order is followed. If
an intersection appears, we must have at least four scan points on
the crossing lines. Denote these points by $p_1, p_2, p_3$ and
$p_4$, as shown in the left of Figure~\ref{fig:BewLe3*C1}. $p_1,
\ldots, p_4$ are located on the paths to or from the essential
edges, or on these essential edges. Without loss of generality, the
essential edges related to $p_1, \ldots, p_4$ lie in clockwise order
on the boundary of $P'$.

The following cases can occur:
\begin{enumerate}
\item There is no scan point between the $p_i$ on the paths.
Then Figure \ref{fig:BewLe3*C1} shows a route that is shorter by
triangle inequality: visit $p_2$ directly after $p_1$, and $p_4$
after $p_3$.

\begin{figure}[h]
\centering
\epsfig{file=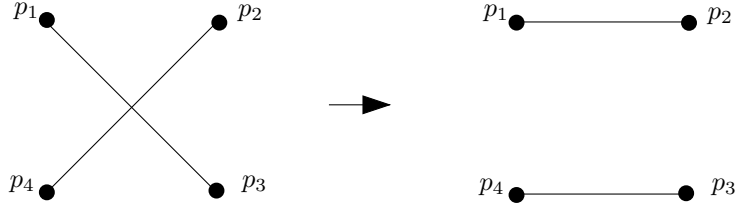}\\
\caption{\label{fig:BewLe3*C1} If there is no scan point between the
$p_i$, the route may be shortened like this.}
\end{figure}

\item If a scan point is located on the intersection point, we need to consider two cases:
\begin{enumerate}
\item Either a scan point on one of the lines established in (1.) is sufficient to see all points,
then the route in (1.) plus this scan point provides lower cost than
the original route.
\item Or a scan point on one of the lines established in (1.) is not sufficient; in that case we
connect two consecutive points by a direct path and use a path via
the intersection scan point for the two other points (or if possible
a path via a point in shorter distance to two of the $p_i$),
cp.~Figure~\ref{fig:BewLe3*Cb}.

\begin{figure}[h]
\centering
\epsfig{file=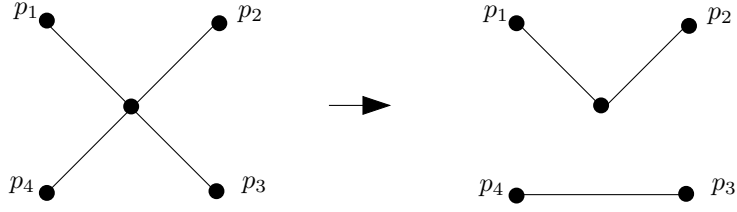}\\
\caption{\label{fig:BewLe3*Cb} If a scan point is located on the
intersection point, but a scan point on one of the lines established
in (1.) is not sufficient, the route may be shortened like this.}
\end{figure}

\end{enumerate}
\item If one of the $p_i$ is the intersection point, we have to consider
the route more closely. For that purpose we mention two properties
of essential extensions in rectilinear polygons, which were stated
by Deng et al.\ \cite{deng98how}.
\begin{proposition}[Proposition 2.2 of Deng et al.\ \cite{deng98how}]\label{exts}\hspace{4cm}
\begin{itemize}
\item[(i)] Two distinct essential extensions are either disjoint or perpendicular to each other.
(Note that the same essential extension may be the extension of two
different sides.)
\item[(ii)] Each essential extension intersects at most two other essential extensions. (If it
intersects two other essential extensions, then these two are
parallel to each other.)
\end{itemize}
\end{proposition}

The general situation is as shown in Figure \ref{fig:BewLe3*AS}.

\begin{figure}[h]
\centering
\epsfig{file=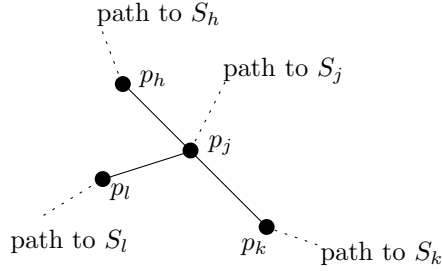}\\
\parbox{13.5cm}{ \caption{\label{fig:BewLe3*AS} The general situation with one of the $p_i$ being the intersection
point. We have $h \neq j, h \neq k, h \neq l, j \neq k, j \neq l$
and $k \neq l$. }}
\end{figure}


The paths leading to (or coming from) the essential extensions may
have length 0, i.e., the corresponding $p_i$ is located on an
essential extension.
\begin{itemize}
\item[(a)]All paths have length $0$:\\
Thus, all $p_i$ are located on essential extensions. The essential
extension on which $p_j$ lies must run along
$\overline{p_{h}p_{k}}$, because (i) if it cuts
$\overline{p_{h}p_{k}}$, $p_h$ or $p_k$ would lie in $P\setminus
P'$, and (ii) the essential extension may not be shorter than
$\overline{p_{h}p_{k}}$, as running along $\overline{p_{h}p_{j}}$,
$\overline{p_{j}p_{k}}$ (independent of direction) would not be
possible otherwise. As a result, $\overline{p_{h}p_{k}}$ is
completely located on the essential extension. This essential
extension is intersected by at most two other essential extensions
(see Proposition \ref{exts}(ii)), which may not intersect between
$p_h$ and $p_k$ (because $p_i$ in $P\setminus P'$.)

In the following we distinguish if $p_1$ , $p_4$, or one of the
points $p_2$ and $p_3$ is the intersection point.

If $p_1$ is the intersection point, not both $p_2$ lying before
$p_4$ in clockwise order and the starting point being located
between $p_4$ and $p_1$, which lie on the same extension, may be
true, leading to a contradiction. The same argument holds for $p_4$
lying on $\overline{p_{1}p_{3}}$.

Otherwise, i.e., if $p_2$ or $p_3$ is the intersection point, these
are the only points where the direction is changed, i.e., no
``real'' intersection appears, and this does not touch the
considered order, see Figure \ref{fig:p2oderp3}, but the routes may
be shortened.

\begin{figure}[h]
\centering
\epsfig{file=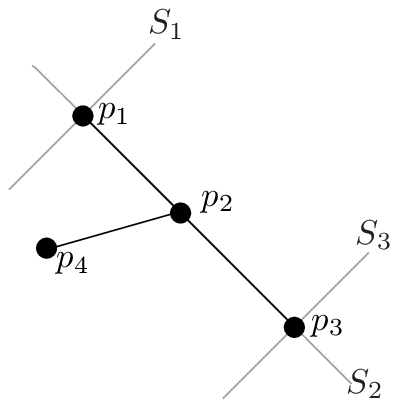}\hspace{1cm}\epsfig{file=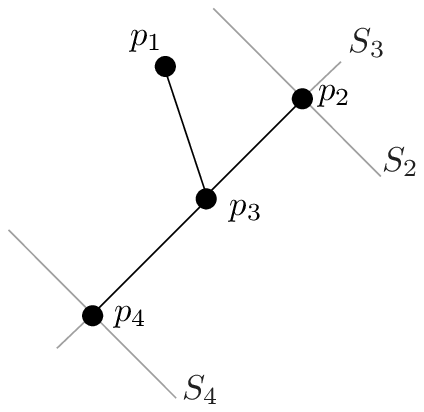}\\
\parbox{12cm}{ \caption{\label{fig:p2oderp3} Left: $p_2$ as "intersection" point, in case of path length = 0. Right:
$p_3$ as "intersection" point, in case of path length = 0. }}
\end{figure}

\item[(b)] At least one path has positive length:\\
\begin{itemize}
\item If $p_j = p_2$ or $p_j = p_3$, the route may turn once or thrice.\\
When the robot turns only once, $p_j$ must be located on an
essential extension, as a path to $S_j$ would cause another turn.
Thus, $p_h$ and $p_k$ must be located on the same extension (see
above), and only a path to $S_l$ with positive length is possible.
This $p_j$ does not influence the requested order, i.e., it is not a
``real'' intersection point.

If the robot turns thrice at $p_j$ (if the path to $S_j$ has
positive length), a loop occurs; then the robot may traverse this
loop in a way that observes the given order.

\item If $p_j = p_1$, the route may only use this point thrice.\\
If the route does not turn thrice at $p_1$, it must start there, as
otherwise $\overline{p_{2}p_{4}}$ is an essential extension, and so
$p_3$ may not lie on a shortest tour, as it would be located in
$P\setminus P'$.

If the route turns thrice, the above loop argument holds.

\item $p_j = p_4$ is analogous to $p_j = p_1$ (with ends instead of
starts).
\end{itemize}
\end{itemize}
\end{enumerate}
\end{proof}

\section{Strategy Aspects}
\label{sec:strategy}

Here we give an overview of the structure of our method; first we
give some basic tools (binary search and turn adjustments in
\ref{subsec:binary} and \ref{subsec:turn}, respectively), followed
by high-level case distinctions in the strategy
(Sections~\ref{subsec:decisions}, \ref{subsec:wtg},
\ref{subsec:dtam} and~\ref{subsec:htm}) and a detailed pseudocode
for our strategy SCANSEARCH (Section~\ref{subsec:strategy}); for
easier reference in checking technical details, we give line
numbers. Finally, in Section~\ref{subsec:scansearch} we consider the
competitive ratio of our strategy.


\subsection{Binary Search in the Strategy}\label{subsec:binary}
As we will often run beyond a point up to which everything is
already known, our strategy may force the robot to pass some
non-visible regions, which are explored with a binary search
strategy. The demand that each edge of the polygon needs to be fully
visible from one scan point implies that the robot needs at most $k$
searches ($2k$ if we have NVRs on both sides, see Lemma \ref{k-2k})
if the optimum uses $k$ scans. This yields a benchmark for computing
the cost of the optimum to determine an upper bound on the
competitive ratio.

If we are confronted with one or more non-visible regions lying in
an area already passed, the maximum width of the passed area is an
upper bound for each possible NVR. Consider a maximum width $w$, and
a minimum side length $a$. Then the binary search can be terminated
after at most $j^{*} = {\log(\frac{w}{a})}$ steps; the total cost is

\begin{equation}
\sum_{j=1}^{\lceil j^{*} \rceil}w \cdot 2^{-j} + \sum_{j=1}^{\lceil
j^{*} \rceil}1 \le w - \frac{a}{2} + {\log(\frac{w}{a})} +1
\end{equation}

The second sum results from the scans after each move. If we have
more than one NVR, we begin with the easternmost, i.e., the one that
is closest to the starting point of the move. This may split some
NVRs into several NVRs, which are all identified.

\begin{lemma}\label{k-2k}If the optimum needs $k$ scans in an interval (of width B), the
robot needs at most
\begin{tabbing}
(ii) \=$2k$ binary searches\kill

(i) \>$2k$ binary searches if the NVRs may appear on two sides,\\
(ii) \>otherwise $k$ binary searches (with an upper bound given by
the above value).

\end{tabbing}

\end{lemma}
\begin{proof}
If the optimum needs $k$ scans (in the interval of width $B$), we
will have $k$ stairs or niches. 
These will only be visible from the running line if a scan is taken
in the integer visibility polygon of the northernmost horizontal
edge, as the boundary runs rectilinear, cp.\ Figure \ref{Bewk2k}.
Each of these northernmost horizontal edges lies inside a NVR. These
are identified by the robot and each NVR has a width less than or
equal to the maximum width. Thus, we need at most a binary search
over $mw$ for each of them.\par

\begin{figure}[h]
\centering
\epsfig{file=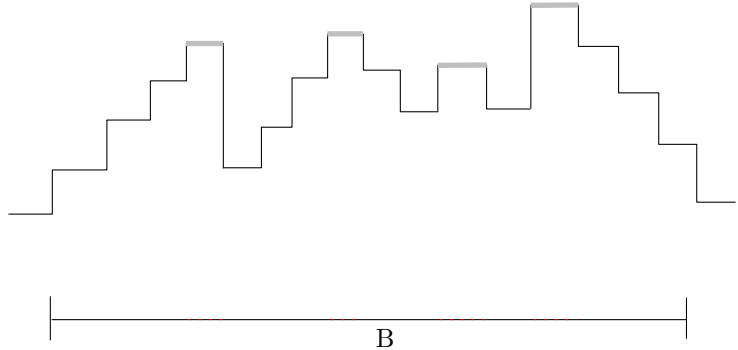}\\
\caption{\label{Bewk2k} An interval of width B. Scans are required
in the integer visibility polygons of the light gray edges.}
\end{figure}

If the non-visible regions appear on two sides (i), we will need
$2k$ binary searches---as a situation like in Figure \ref{2knotw}
may occur. That is, with our strategy the robot distinguishes the
NVRs, reaches one of the dark gray positions, where one, but not
both, of the non-visible regions become visible. Thus, the robot
will start another binary search. Consequently, if the optimum takes
$k$ scans, the robot will need at most twice as many binary
searches.\par

\begin{figure}[h]
\centering
\epsfig{file=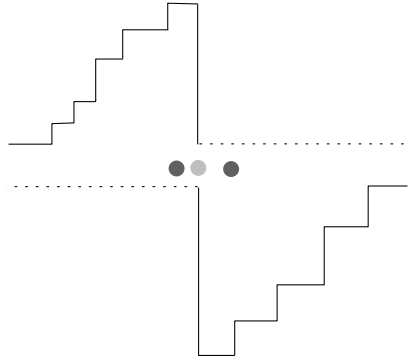}\\
\parbox{12cm}{ \caption{\label{2knotw} Worst case for the scan points of the optimum (light gray) and
our strategy (dark gray) if the non-visible regions appear on two
sides.}}
\end{figure}

\end{proof}

\subsection{Turn adjustments}\label{subsec:turn}
In some cases our robot needs to make a turn, but we do not know
where the optimum turns (see Figure~\ref{fig:stepchoice}(d)). We
handle this uncertainty by considering the maximum corridor possible
for turning, move in its center and make an adjustment to the new
center whenever a width reduction occurs. Note that we do not make
an adjustment when a width increases. When the robot is supposed to
make another turn, we adjust to the best possible new position
within the corridor. This procedure will be called \textit{turn
adjustment} in the following. The adaptions we apply are described
in the proofs of the according lemmas.

Turning in an optimal solution may be the result of a regular turn
(Lemma~\ref{casea}), a corridor becoming visible in an NVR, in which
case it can be reached by an axis-parallel motion
(Lemma~\ref{caseb}) or in case the south and eastern boundary are
closed (Lemma~\ref{casec}.) In all these cases, we need to make sure
that our strategy does not incur too high marginal costs compared to
the optimum. We assume that each corridor in the polygon has a
minimum corridor width and refer to it as $a_k$. With this
assumption we know up to which bound we may have to reduce the step
length in a binary search.

\begin{lemma}\label{casea}
Suppose that in an optimal tour, the optimum turns earlier than we
choose to in our strategy. Then the marginal cost of the
corresponding turn adjustment remains within a factor of $O(\log A)$
of the optimum.
\end{lemma}

\begin{figure}[h]
\centering
\epsfig{file=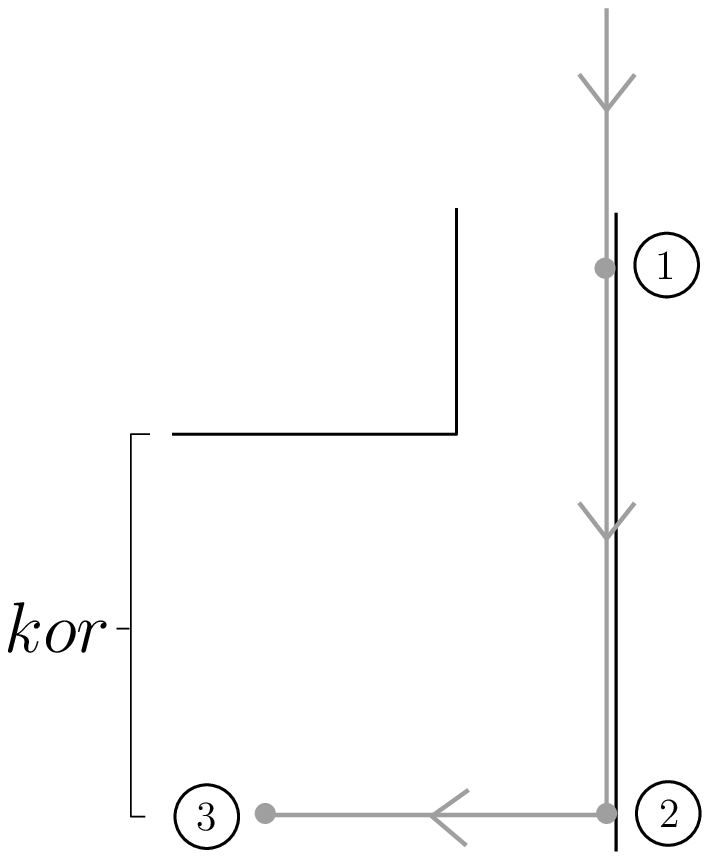, height=3cm}\hspace{0.7cm}\epsfig{file= 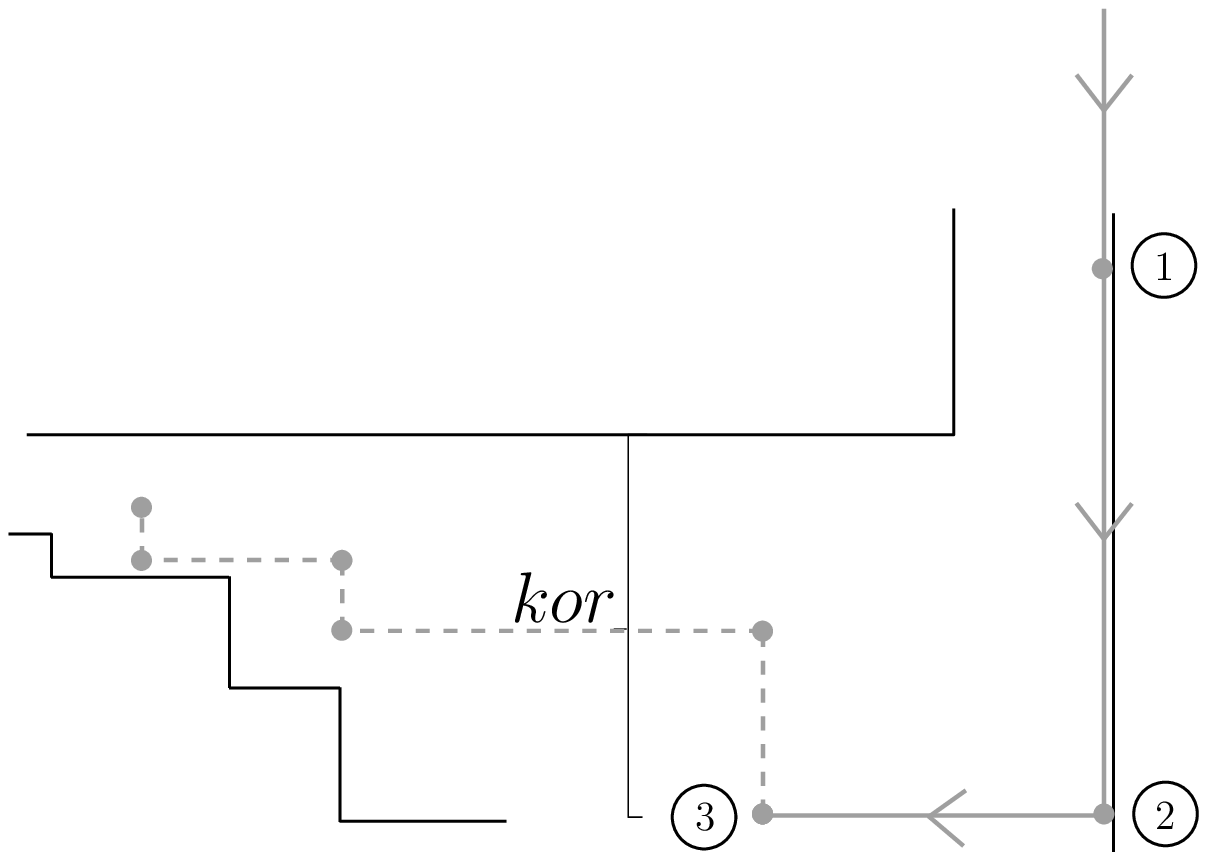, height = 3cm }\hspace{0.7cm} \epsfig{file=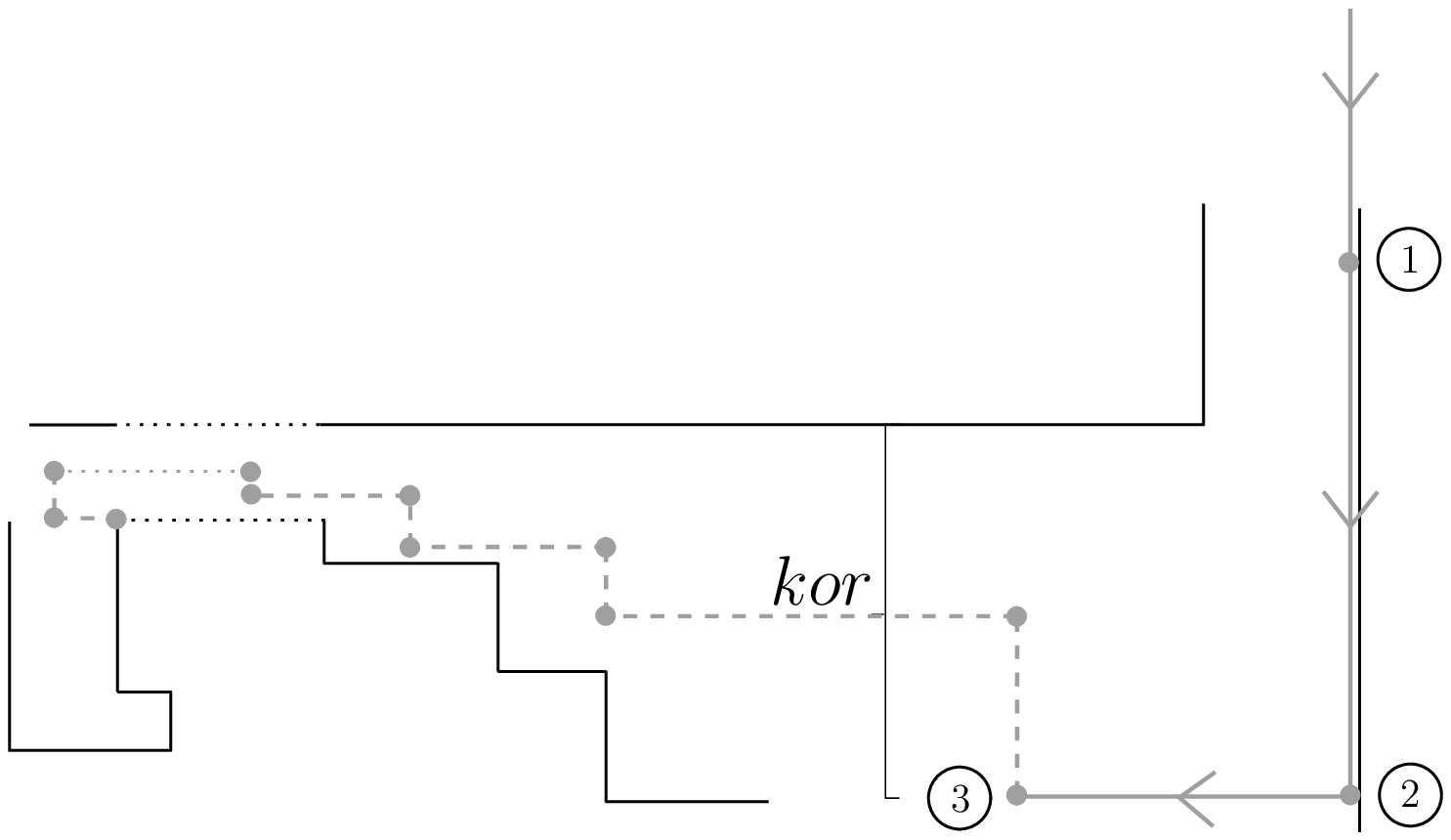,height=3cm} \\
 \caption{\label{corridor} An interval of width {\em kor},
in which the optimum could turn earlier. The numbers indicate the
order in which the scans are performed.}
\end{figure}

\begin{proof}
Consider the width $kor$ of the interval in which the optimum may
turn earlier, see Figure \ref{corridor} left. The interval of width
\textit{kor} in which an axis-parallel movement is possible may
become narrower because of the boundary (Figure~\ref{corridor}
middle). If this keeps the robot from running axis-parallel, the
robot runs vertically to the center of the remaining interval, etc.
The total cost for these adjustments does not exceed the cost of a
binary search in an interval of width {\em kor}. If during the
search of the non-visible regions we realize that we need to deviate
to the south or the north from the horizontal line, i.e., if we find
a corridor, we adapt to the best possible position (cost of $kor/2 +
1$). That is, we adapt to the best height and add a step to the
easternmost part of the corridor, if this lies to the south
(Figure~\ref{corridor} right).
\end{proof}

\begin{lemma}\label{caseb}
Suppose that during the course of the exploration, a corridor is
discovered inside of a non-visible region, forcing any optimal
solution to make a turn. Then the marginal cost of our strategy
remains within a factor of $O(\log A)$ of the optimum.
\end{lemma}

\begin{proof}
When we discover a corridor, the NVR does not consist of stairs or
niches. If the NVR lies south, we look for the first possible
eastern corridor, otherwise for the first possible western corridor.
The width of the corridor cannot exceed the distance that we covered
beyond the extension $E$, and we take this distance as {\em kor}. If
in the following it is not possible to continue running vertically,
the robot runs horizontally to the center of the narrower interval,
etc. The costs are estimated by a binary search, and the adjustments
are done analogously.
\end{proof}

\begin{lemma}\label{casec}
Suppose that while a planned axis-parallel move is not possible
without a change of direction and the boundary is closed to the
south, i.e., no unseen corridor lies to the south, a corridor is
discovered that may lie either in the western or the northern area.
Then the marginal cost of our strategy remains within a factor of
$O(\log A)$ of the optimum.
\end{lemma}

\begin{figure}[h]
\centering
\epsfig{file=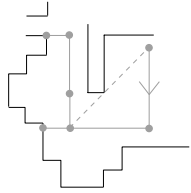, height=2.5cm} \\
 \caption{\label{corridorIII} The boundary is closed to the south. Thus, the corridor may either lie to the west or the north. We allow for adjustments in both areas.}
\end{figure}

\begin{proof}
We proceed analogously to the previous argument. Note that the
adjustments can happen twice, first in the western, then in the
northern area, see Figure~\ref{corridorIII}. Thus, we need two times
the upper bound of the binary search in $kor$, $kor/2$ and $1$.
\end{proof}

\subsection{High-Level Decisions within Strategy SCANSEARCH}
\label{subsec:decisions} Just like in the GREEDY-ONLINE strategy by
Deng et al., we start with identifying the next extension $E$.
Without loss of generality, let the known parts of the boundary run
north-south and east-west, and the next extension run north-south,
either defined by the bound of the contiguous visible part of the
boundary, $f$, or by a sight-blocking corner, $b$. The boundary is
in clockwise order completely visible up to the extension. The
minimum side length of the polygon $P$ is given and denoted by $a$.

\noindent When we want to move we have to
\begin{itemize}
\item determine which point we head for and
\item decide how to move there.
\end{itemize}

\subsection{Where to Go}\label{subsec:wtg} To determine the next scan point we
first choose a reference point---which, in turn, depends on the next
extension in clockwise order.

\begin{case_d}[Axis-Parallel Movement]\label{one}We may \hspace*{5cm}
\begin{itemize}
\item[[A.\hspace{-.2cm}]] reach the next extension, using an axis-parallel move without a turn
or (line 2)
\item[[B.\hspace{-.2cm}]] not reach the next extension axis-parallel without a
change of direction (line 32).
\end{itemize}
\end{case_d}

\noindent Our first toehold is the ($L_2$-)distance $e$ to $E$---a
small distance does not allow for a scan on $E$ and, thus, we will
walk beyond it (\textit{extension case}). $e$ being large enough
results in exploring the area up to $E$ (\textit{interval case}).
Here, ``large'' and ``small'' depend on the subcases.
\begin{case_d}[Interval and Extension Case]\label{two}\hspace*{5cm}
\begin{itemize}
\item Case [A.]: for $e \geq 2a+1$ (large) we are in interval case (line 3), else in
extension case (line 24).
\item Case [B.]: for $e \geq a+1$ (large) we are in interval case (line 33), else in
extension case (line 90).
\end{itemize}
\end{case_d}
\noindent Whereas we get a reference point for the extension case,
we need to consider other points for the interval case:

\noindent {\bf Let us first assume to be in case [A.].}
\begin{case_d}[interval case in [A.\mbox{]}]\label{three}
The next reference point is
\begin{itemize}
\item either the perpendicular of the next counterclockwise corner to
the shortest path to $E$ ($d_i$ being the distance to this point)
(line 4 et seqq.) or
\item the point on $E$ within distance $e$, if no NVRs appear on the
counterclockwise side (line 9).
\end{itemize}
\end{case_d}
\noindent {\bf So, let us now assume to be in case [B.].}
\begin{case_d}\label{four}
In case an axis-parallel move to $E$ is not possible without a
change of direction and $e \ge a+1$, there may be
\begin{tabbing}
($\alpha$) \= bla \kill ($\alpha$) \> no non-visible regions up to
the sight-blocking corner (line 34 et seqq.),\\
($\beta$) \> or non-visible regions up to the sight-blocking corner
(line 66 et seqq.).
\end{tabbing}
\end{case_d}
For ($\alpha$) our point of reference is the sight-blocking corner,
let $b_i$ be the current distance to this corner.

For ($\beta$) we consider the intersection points of the line
between the start position of this case and the sight blocking
corner and the extension of the invisible adjacent edge of the next
corner on the east-west boundary as well as the extension of the
invisible adjacent edge of the next corner on the north-south
boundary. Our point of reference is the intersection point with
smaller distance, $m_i$, to the current position. (Look at
Figure~\ref{Beispiel} for an example of these cases.)

\subsection{Decisions That Affect a Move}\label{subsec:dtam}

So far, we only decided on the reference point for our next move.
Now we describe how to determine the point we head for, handle
events that occur during movements, and how to perform a move.

In general:
\begin{case_d}[Planned Distance]\label{five}
When we face a large distance to the next reference point, we simply
go there. When we face a {\textit small} distance, $r$, to the next
reference point, we plan to cover a distance of $2r+1$.
\end{case_d}

\begin{figure}[h]
\centering \epsfig{file=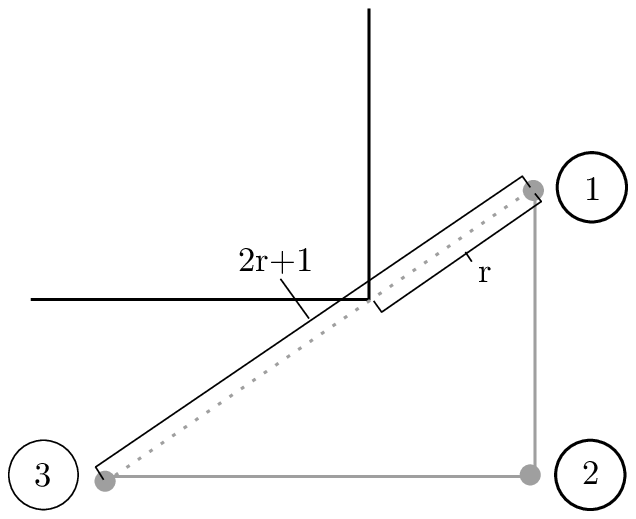, height =3.0cm} \hspace*{3cm}
\epsfig{file=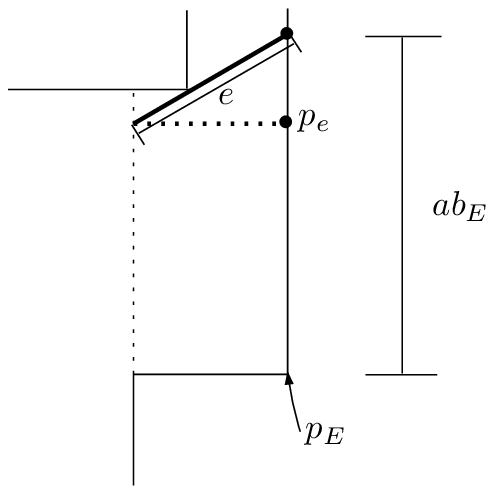, height=3cm}\\
\caption{\label{abEetc} Left: covering a distance of $2r+1$ along
the straight connection in moving axis parallel (the numbers
indicate the order in which the scans are performed); right: an
example for $p_e, p_E$ and $ab_E$.}
\end{figure}

\begin{case_d}[Crossing a Given Extension]\label{six}
Let $r$ be the distance to a given reference point. If we plan to
cover a distance of $2r+1$ we may
\begin{tabbing}
(ii) \=$2k$ binary searches\kill (i)  \>either not be able to cover
the total planned length
because of the boundary,\\
(ii) \>or be able to cover the distance of $2r+1$.
\end{tabbing}
\end{case_d}
Walking a distance of $2r+1$ implies following the axis-parallel
line if this is possible ([A.]). Otherwise ([B.]) we end up within a
distance of $2r+1$ along the straight connection and go there by
walking in an axis-parallel fashion, see Figure \ref{abEetc} (left).

\begin{case_d}[Line Creation]\label{seven}
Whenever we are in case (ii) of Case \ref{six}, we draw an imaginary
line parallel to the extension running through the current position.
Then we observe whether the entire boundary on the opposite side of
this line is visible.
\begin{tabbing}
(ii) \= bla \kill
(I) \> If this is the case we say that the line creation is \emph{positive} (lines 21, 31, 52, 63, 78, 88 and 102)\\
(II) \> Otherwise we refer to it as a \emph{negative} line creation
(see for example Figure~\ref{Beispiel}) (lines 17, 30, 51,\\
\> 62, 77, 87 and 101).
\end{tabbing}
\end{case_d}

\noindent In case of a positive line creation (I), which implies
that we do not have to keep on searching in the area behind the
line, we move back to $E$ and start searching for a NVR. Otherwise
we apply binary search and
enter corridors in southern NVRs.\\

\noindent Moreover, for case [B.], we have:
\begin{case_d} [Covering the Total Planned Length Is Not Possible]
\label{eight} If we meet the preconditions of Case \ref{six}(i) and
an axis-parallel move to $E$ is not possible without a change of
direction, the boundary is
\begin{tabbing}
(a) \= bla \kill (a) \> either not closed south of the path (lines 44, 59, 74, 84 and 98)\\
(b) \> or closed south of the path (lines 47, 60, 75, 85 and 99).
\end{tabbing}
\end{case_d}

\subsection{How to Move}\label{subsec:htm}
 \noindent The further movements and
actions depend on the distances in cases
[A.] and [B.]:\\
 {\bf [A.]:}
 \begin{itemize}
 \item If the distance to the next reference point is big enough, we
 cover this distance.
 \item Otherwise we walk beyond this point (Case Distinction \ref{six}).\\
Case \ref{six}(i) results in moving as far as possible, moving back
to $E$, applying binary search for NVRs (up to $E$ on one side,
beyond $E$ on both sides) and using a corridor whenever we find one
(with turn
adjustments).\\
Case \ref{six}(ii) is the precondition of Case Distinction
\ref{seven}.
 \end{itemize}
{\bf [B.]:}\\ \noindent In the extension case without the
possibility to reach $E$ axis-parallel without a change of direction
(l 90 et seqq.) we refer to the point in which the axis-parallel
move to $E$ changes the direction as $p_e$. $p_E$ is the next corner
on the boundary in clockwise order, and $ab_E$ is the distance from
the current starting point to $p_E$, see Figure \ref{abEetc}
(right). With $ab_E$ we have again the critical distance.

\begin{case_d}\label{nine}
If we face the extension case without the possibility to reach $E$
axis-parallel without a change of direction, we distinguish two
possible motions:
\begin{itemize}
\item If $ab_E$ is large ($ab_E > 2a+3$) we cover a distance of $2e+1$
along the straight connection in moving axis-parallel. (The distance
to $p_E$ allows us to do so on the first axis-parallel line.) This
results again in the basic case distinctions \ref{six}, \ref{seven}
and \ref{eight}.
\item For $ab_E \le 2a+3$ (small) we move to $E$ via $p_E$ and---if
necessary---apply binary search and make turn adjustments.
\end{itemize}
\end{case_d}

\vspace*{-0.2cm} \noindent The interval case without the possibility
of reaching $E$ axis-parallel without a change of direction (line 33
et seqq.) requires some more case analysis.

\begin{figure}[h]
\centering
\epsfig{file=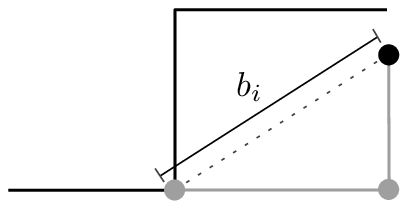, height=1.5cm}\hspace{1cm}\epsfig{file=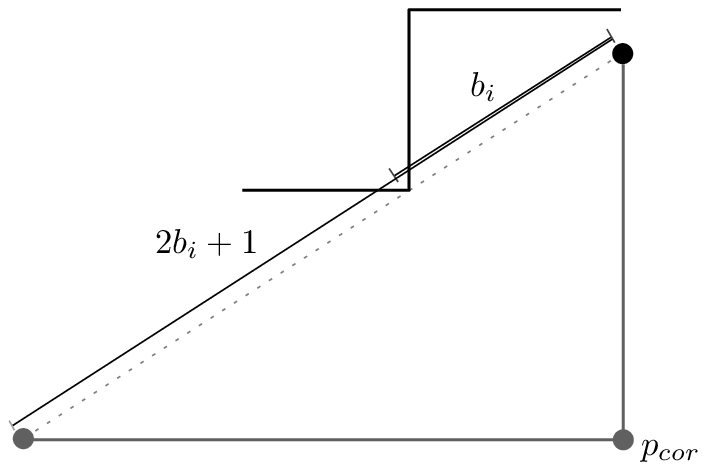, height=3cm}\hspace{1cm}\epsfig{file=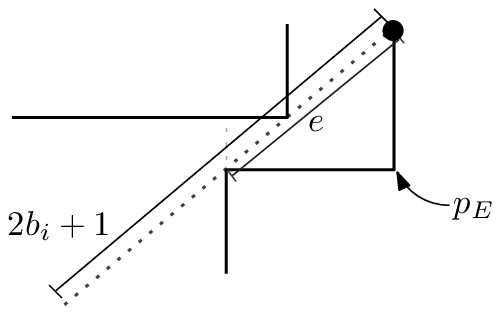, height=3cm}\\
{ \caption{\label{StrFig} Left:If $b_i > 2a+1$, the robot moves
axis-parallel to the corner. Middle: If $b_i \le 2a+1$, the robot
moves axis-parallel to a point in distance $2d_i+1$, and in doing so
it visits $p_{cor}$. Right: $p_E$ is the point in which the
clockwise boundary bends to $E$.}}
\end{figure}

\noindent ($\alpha$): Let $b_i$ be the distance to the
sight-blocking corner, which is our point of reference.
\begin{itemize}
\item Thus, if $b_i$ is larger than
$2a+1$, we walk to the sight-blocking corner. As always when
axis-parallel moves without a turn are not possible, we cover this
distance in an axis-parallel fashion (line 35), cp.~Figure~\ref{StrFig} (left).\\
\item  \begin{case_d}\vspace*{-0.75cm}\label{ten}If $b_i < 2a+1$ holds:
\begin{itemize}
\item[($\mathcal{A}$)]  Either we may be able to cover a distance of
$2b_i+1$ and run beyond $E$ on the second axis-parallel line (after
the change of direction), which results in the case distinction (and
movements) of Case distinctions \ref{six}, \ref{seven} and
\ref{eight} (line 41 et seqq.), cp.~Figure~\ref{StrFig} (middl)e.
\item[($\mathcal{B}$)]  Or the point where we would have to change our
direction ($p_{cor}$) may not lie inside the polygon (line 53). So
we walk to $p_E$ (see Figure \ref{abEetc} (right) for the definition
and Figure~\ref{StrFig} (right) for the actual move) and then
axis-parallel to the straight connection. In doing so we
\begin{tabbing} $\bullet$ \= bla \kill
$\bullet$ \> either do not run beyond $E$ (line 54)\\
$\bullet$ \> or would run beyond $E$ (line 57), resulting again in
the case distinctions \ref{six}, \ref{seven} and
\ref{eight}.\end{tabbing}
\item[($\mathcal{C}$)]  Otherwise turn adjustments are used (line 64).
\end{itemize}
\end{case_d}
\end{itemize}

\noindent ($\beta$): With NVRs appearing up to the sight-blocking
corner \textit{($\beta$)} we consider other points of reference, but
the structure is the same as in \textit{($\alpha$)}. The critical
distance $m_i$ is the shortest distance to the intersection point of
the straight connection to the sight-blocking corner and the
extension of one side of an NVR. Moreover, the point that is
equivalent to $p_{cor}$ is called $p_{m}$ (line 66 et seqq.).

For $a > 1$ we use a similar strategy (line 103 et seqq.). Because
scans are taken whenever a distance of $a$ is covered, NVRs are
explored while passing and corridors are identified immediately.

While exploring the polygon, we make sure that in clockwise order
all parts of the polygon are visible after having been passed, i.e.,
we make sure that we see everything a watchman with continuous
vision would see when walking along the basic path. Areas that are
not visible define an extension in the remaining part of the
algorithm, and we always use the next clockwise corridor. Moreover,
we return to the starting point as soon as we have seen all sides of
the boundary.

\subsection{The strategy SCANSEARCH}\label{subsec:strategy}

\SetKwBlock{ABAnfang}{}{} \SetKwData{A}{[A.]}
\SetKwData{B}{[B.]}\SetKwData{Stern}{[*]}\SetKwData{eins}{[1.]}
\SetKwData{zwei}{[2.]}\SetKwData{drei}{[3.]}\SetKwData{ieins}{(i)}
\SetKwData{izwei}{(ii)}\SetKwData{scrA}{($\mathcal{A}$)}\SetKwData{scrB}{($\mathcal{B}$)}
\SetKwData{scrC}{($\mathcal{C}$)}\SetKwData{Ieins}{(I)}
\SetKwData{Izwei}{(II)}\SetKwData{Alpha}{($\alpha$)}\SetKwData{Beta}{($\beta$)}
\SetKwData{kleinA}{(a)} \SetKwData{kleinB}{(b)}
\SetKwData{Punkt}{$\bullet$ }
\begin{algorithm} [H]
\caption{The strategy SCANSEARCH} \small \KwIn{A starting position
inside an unknown rectilinear polygon $P$, its minimum side length
$a$, its minimum corridor width $a_k$.} \KwOut{A route along which
the whole polygon becomes visible for a robot with discrete vision,
such that each edge is fully visible from some scan point.}

We identify the next extension in analogy to the GREEDY-ONLINE
algorithm of Deng et al.\ \cite{deng98how}, i.e., we update $C,f$
and $M$ whenever changes occur.

If $f$ is a reflex corner, let $E = Ext(F(f))$. Otherwise let $b$ be
the blocking corner when $f^{-}$ was in view, and let $E
=Ext(B(b))$. \\
If it is possible to move axis-parallel to $E$ without a change of
direction, let $e$ be the axis-parallel distance to $E$; otherwise,
let $e$ be the shortest distance to $E$ without a change of
direction.\\
 \nl $a \le 1$:
 \ABAnfang{ \nl \A An axis-parallel move to $E$ is
possible without a turn: \ABAnfang{ \nl \Stern $e \ge 2a+1$:
\textbf{interval case}\ABAnfang{\nl \eins If $d_i > 2a+1$, move to
the perpendicular of the corner.} \vspace*{-0.85cm}\ABAnfang{\nl
\zwei If $d_i \le 2a+1$: \;  \nl if $d_i
> a$: cover a distance of $2d_i+1$\;  \nl if $d_i \le a$: cover a
distance of $2a+1$\; \nl apply binary search if necessary, i.e., if
non-visible regions appear.} \vspace*{-0.9cm}\ABAnfang{\nl \drei If
no corner appears on the counterclockwise side, move directly to
$E$.}\vspace*{-0.4cm}\nl If we run beyond $E$ with a step of length
$2d_i+1$/$2a+1$: \ABAnfang{\nl
\ieins If we do not cover the total distance, because of the boundary:\\
\nl Run as far as possible, go back to $E$, move back in steps of
length 1, apply binary search \nl for NVRs (on the counterclockwise
side till $E$, on both sides beyond $E$); if a corridor is \nl
identified, use it and make turn adjustments. (If a critical
extension is found, search only on \nl the opposite
side.)}\vspace*{-0.9cm} \ABAnfang{\nl  \izwei If we may cover the
total distance of $2d_i+1$/$2a+1$: \ABAnfang{\nl  \Ieins negative
line creation: \\ \nl Apply binary search; if a corridor is
discovered inside an NVR, use it and make turn \nl adjustments.
(Because the line creation is negative, only corridors in southern
non-visible \nl regions are used.)} \vspace*{-0.85cm}\ABAnfang{\nl
\Izwei positive line creation:\\\nl  Go back to $E$, move back in
steps of length 1, apply binary search and search for a corridor \nl
and the critical extension, making turn
adjustments.}}}\vspace*{-1.3cm}\ABAnfang{\nl \Stern $e < 2a+1$:
\textbf{extension case} \\ \nl Consider running a distance of
$2e+1$. \ABAnfang{\nl  \ieins If it is not possible to run a
distance of $2e+1$: Run as far as possible, go back to $E$, move
back in \nl steps of length 1, apply binary search for NVRs and, if
a corridor is identified, use it and make turn \nl
adjustments.}\vspace*{-0.9cm}\ABAnfang{\nl  \izwei If we may cover
the total distance of $2e+1$:\ABAnfang{ \nl \Ieins negative line
creation.}\vspace*{-0.85cm}\ABAnfang{\nl  \Izwei positive line
creation.}}}}
\end{algorithm}
\newpage

\begin{algorithm}[H]\small 
\ABAnfang{\nl \B An axis-parallel move to $E$ is not possible
without a change of direction: \ABAnfang{\nl \Stern $e \ge a+1$:
\textbf{interval case}\\ \nl \Alpha No non-visible region up to the
sight-blocking corner \ABAnfang{\nl \eins If $b_i > 2a+1$, move
axis-parallel to the corner, see Figure \ref{StrFig}, left.}
\vspace*{-0.9cm}\ABAnfang{\nl \zwei If $b_i \le 2a+1$, cover a
distance of $2b_i+1$ along the straight connection in moving
axis-parallel \nl and visiting $p_{cor}$, see Figure \ref{StrFig},
middle. If necessary, apply binary search. Apply the binary search
\nl on the first axis-parallel line before leaving $p_{cor}$ in a
right angle to this line and \nl apply the binary search on the
second axis-parallel afterwards.
\\\nl  Now distinguish the following. \ABAnfang{ \nl \scrA If we run
beyond $E$ (on the second axis-parallel line): \ABAnfang{\nl \ieins
If it is not possible to cover the total planned length, let
$\alpha_i$ be the distance to this \nl boundary along the straight
connection, and \ABAnfang{\nl \kleinA if the boundary is not closed
south of the path, i.e., the clockwise exploration of \nl the
polygon continues south of the second axis-parallel line, then move
as far as possible, apply \nl binary search on both axis-parallel
lines and make turn adjustments.}\vspace*{-0.9cm}\ABAnfang{\nl
\kleinB if the boundary is closed south of the path, then move as
far as possible, apply \nl binary search on both axis-parallel lines
and apply turn adjustments, if necessary twice, as we are \nl in the
last case of the turn adjustments.}} \vspace*{-1.3cm}\ABAnfang{\nl
\izwei If it is possible to cover the total planned length, we
distinguish: \ABAnfang{\nl \Ieins negative line
creation.}\vspace*{-0.85cm}\ABAnfang{\nl \Izwei positive line
creation.}}}

\vspace*{-1.3cm}\ABAnfang{\nl \scrB If it is not possible to run via
$p_{cor}$ (as the boundary blocks us from doing so) and
\ABAnfang{\nl  \Punkt if we do not run beyond $E$ in doing so: Run
to $p_E$ (see Figure \ref{StrFig}, right) and then \nl axis-parallel
to the straight connection. If necessary, apply binary search and
make turn \nl adjustments.} \vspace*{-0.9cm}\ABAnfang{\nl  \Punkt if
we would run beyond $E$: \ABAnfang{\nl \ieins If it is not possible
to cover the total planned length: \ABAnfang{ \nl \kleinA If the
boundary is not closed south of the path.}
\vspace*{-0.85cm}\ABAnfang{\nl \kleinB If the boundary is closed
south of the path.}} \vspace*{-1.3cm}\ABAnfang{\nl \izwei If it is
possible to cover the total planned length, we distinguish:
\ABAnfang{ \nl \Ieins negative line
creation.}\vspace*{-0.85cm}\ABAnfang{\nl \Izwei positive line
creation. }}}}

\vspace*{-1.3cm}\ABAnfang{\nl \scrC neither ($\mathcal{A}$) nor
($\mathcal{B}$) is true, make turn adjustments (like in the first
case of the turn \nl adjustments).

}}}
}
\end{algorithm}

\newpage
\begin{algorithm}[H]\small\ABAnfang{ \ABAnfang{\nl \Beta Along the boundary up to the
sight-blocking corner occur non-visible regions \ABAnfang{\nl  \eins
If $m_i > 2a+1$, cover a distance of $m_i$ along the straight
connection in moving axis-parallel and \nl visiting $p_m$.}
\vspace*{-0.9cm}\ABAnfang{\nl \zwei If $m_i \le 2a+1$, cover a
distance of $2m_i+1$ along the straight connection in moving \nl
axis-parallel and apply binary search if necessary.}
 \nl In (2.) several cases may occur: \ABAnfang{\nl
\scrA If we run beyond $E$ (on the second axis-parallel line):
\ABAnfang{\nl \ieins If it is not possible to cover the total
planned length: \ABAnfang{\nl \kleinA If the boundary is not closed
south of the path.}\vspace*{-0.85cm}\ABAnfang{\nl \kleinB If the
boundary is closed south of the path.}}
\vspace*{-1.3cm}\ABAnfang{\nl \izwei If it is possible to cover the
total planned length, we distinguish: \ABAnfang{\nl \Ieins negative
line creation.} \vspace*{-0.85cm}\ABAnfang{\nl \Izwei positive line
creation.} }}

\vspace*{-1.3cm}\ABAnfang{\nl \scrB If it is not possible to run via
$p_{m}$ (as the boundary hinders us to do so) and \ABAnfang{\nl
\Punkt if we will not run beyond $E$ in doing so: Run to $p_E$ and
then axis-parallel to the straight \nl connection. If necessary
apply binary search and make turn
adjustments.}\vspace*{-0.9cm}\ABAnfang{\nl \Punkt if we run beyond
$E$: \ABAnfang{\nl \ieins If it is not possible to cover the total
planned length: \ABAnfang{\nl \kleinA If the boundary is not closed
south of the path.} \vspace*{-0.85cm}\ABAnfang{\nl \kleinB If the
boundary is closed south of the path.}}\vspace*{-1.3cm}\ABAnfang{\nl
\izwei If it is possible to cover the total planned length, we
distinguish: \ABAnfang{\nl \Ieins negative line creation.}
\vspace*{-0.85cm}\ABAnfang{\nl \Izwei positive line creation.}}}}

\vspace*{-1.3cm}\ABAnfang{\nl \scrC If neither ($\mathcal{A}$) nor
($\mathcal{B}$) is true, make turn adjustments.}

 }\vspace*{-0.75cm}\ABAnfang{\nl $e < a+1$: \textbf{extension case}
\ABAnfang{\nl \Punkt If $ab_{E} \le 2a+3$, move to $E$ via $p_{E}$.
If necessary apply binary search and turn adjustments of the \nl
first kind. } \vspace*{-0.85cm}\ABAnfang{\nl \Punkt If $ab_{E} >
2a+3$, cover a distance of $2e+1$ along the straight connection in
moving axis-parallel. \nl This is possible as $ab_{E} > 2a+3$.
Several cases may occur when we want to cover a distance of $2e+1$
\nl along the straight connection in moving axis-parallel:
\ABAnfang{\nl \ieins If it is not possible to cover the total
planned length, let $e_b$ be the distance to this boundary \nl along
the straight connection, and \ABAnfang{\nl  \kleinA if the boundary
is not closed south of the path} \vspace*{-0.85cm}\ABAnfang{\nl
\kleinB if the boundary is closed south of the path.}}
\vspace*{-1.3cm}\ABAnfang{\nl \izwei If it is possible to cover the
total planned length, we distinguish: \ABAnfang{ \nl \Ieins negative
line creation.} \vspace*{-0.9cm}\ABAnfang{\nl \Izwei positive line
creation.}

 }}}}

\end{algorithm}
\newpage
\vspace*{0.1cm}
\begin{algorithm}[H]
\small
 \nl $a \le 1$:\\
\nl Identify the next extension and consider the possibility to
reach it in an axis-parallel fashion without a change \nl of
direction, as well as the distinction between the interval and the
extension case. \ABAnfang{ \nl \A An axis-parallel move to $E$ is
possible without a
turn: \ABAnfang{\nl \Stern \textbf{interval case}\\
\nl Move to $E$ and take a scan each time a distance of $a$ is
covered. In addition, scan on $E$ if scanning with \nl distance of
$a$ does not
result in a scan on $E$.}\vspace*{-0.9cm}\ABAnfang{\nl \Stern \textbf{extension case}\\
\nl If possible, cover a distance of $2e+1$; in doing so, take a
scan each time a distance of $a$ is covered. If a \nl corridor is
discovered to the south of the running line, use it. If it is not
possible to cover a distance of \nl $2e+1$, run as far as possible
(taking a scan each time a distance of $a$ is covered), use a
southern \nl corridor, or, if no southern corridor exists, move back
to the clockwise first northern
corridor.}}\vspace*{-1.35cm}\ABAnfang{\nl \B An axis-parallel move
to $E$ is not possible without a change of
direction: \ABAnfang{\nl \Stern \textbf{interval case}\\
\nl Cover a distance of $e$ along the straight connection in moving
axis-parallel, taking a scan whenever a \nl distance of $a$ is
covered
as well as at the turn and on $E$.}\vspace*{-0.9cm}\ABAnfang{\nl \Stern \textbf{extension case}\\
\nl Cover a distance of $2e+1$ along the straight connection in
moving axis-parallel, taking a scan \nl whenever a distance of $a$
is covered, as well as when the direction is changed and when the
distance is \nl covered. If it is not possible to cover the total
planned length, move as far as possible and take the \nl scans in
analogy to the move described above. Use a southern corridor as well
as a western or northern \nl one, when the total possible distance
is covered.}} Move to the easternmost northern NVR with a corridor
if no corridor appears in the other non-visible regions, if $E$ is
passed and if there is no negative line creation. If everything is
visible between the beginning and the end of the current case, stop
applying binary search, the steps of length 1 etc., and continue
with identifying the next extension.
\end{algorithm}

\vspace*{1.5cm}

\subsection{{\it Correctness and Competitive Ratio  }}
\label{subsec:scansearch}
\begin{theorem}
A simple rectilinear polygon allows an $O(\log A)$-competitive
strategy for the OWPDV ({\textit Online Watchman Problem with
Discrete Vision}), provided each edge needs to be fully visible from
some scan point.
\end{theorem}

\begin{proof}
As noted in Lemmas~\ref{casea}, \ref{caseb}, \ref{casec}, the
marginal cost of our strategy in making turns does not exceed the
optimal marginal cost by more than a factor of $O(\log A)$.

In all cases of the strategy we can relate our cost to a minimum
cost of the optimum (which has to cover a certain distance or has to
cover a certain distance and needs a certain number of scans in the
interval covered). Thus, in each case we are able to limit the
competitive ratio.

The estimate for the competitive ratio is computed from the upper
bounds for the competitive ratio in the different cases:

For computing these estimates, we compare the numerous cases of
strategy SCANSEARCH with the optimum, resulting in the values listed
in Tables \ref{cs1} and \ref{cs2}; a detailed verification is
straightforward, but tedious and left to the reader. Several of
these bounds are dominated, e.g., the value for $k = 0$ is less than
the value for $k
> 0$ in the same case. These dominating values are printed bold,
and, for dominated values with $k
> 0$, the dominating term is labeled in parentheses.
\begin{table}[h]
\centering
\begin{tabular}{|c||c|l|}\hline
$c \le$& $k = 0$ & $k > 0$\\\hline\hline $A.,$ & $\scriptstyle 2$&
\\\cline{2-3} $e \ge 2a+1$ & $\scriptstyle 2$
& \textbf{(1):
$\scriptstyle\frac{3}{2}a+5+\frac{ln\left(\frac{a+1}{a}\right)}{ln(2)}$}\\\cline{2-3}
&$\scriptstyle 2$ &
$\scriptstyle\frac{3}{2}a+5+\frac{ln\left(\frac{a+1}{a}\right)}{ln(2)}$
(1) \\\cline{2-3} &
$\scriptstyle\frac{9}{2}a-\frac{a_k}{2}+22+\frac{ln\left(\frac{2(a+1)}{a_k}\right)}{ln(2)}+\frac{ln\left(\frac{1}{a}\right)}{ln(2)}$
& \textbf{(2):$
\scriptstyle\frac{27}{2}a-\frac{a_k}{2}+29+\frac{ln\left(\frac{1}{a}\right)}{ln(2)}+\frac{ln\left(\frac{2(a+1)}{a_k}\right)}{ln(2)}+2\frac{ln\left(\frac{4a+3}{a}\right)}{ln(2)}$}\\\cline{2-3}
&$\scriptstyle
5a-\frac{a_k}{2}+10+\frac{ln\left(\frac{2(a+1)}{a_k}\right)}{ln(2)}$
& $\scriptstyle
14a-\frac{a_k}{2}+18+\frac{ln\left(\frac{2(a+1)}{a_k}\right)}{ln(2)}+2\frac{ln\left(\frac{4a+3}{a}\right)}{ln(2)}$
(5)\\\hline $A.,$ & $\scriptstyle 2$ & \textbf{(3):
$\scriptstyle7a+9+2\frac{ln\left(\frac{4a+3}{a}\right)}{ln(2)}$}\\\cline{2-3}
$ e < 2a+1$ & $\scriptstyle
14-\frac{a}{2}+\frac{ln\left(\frac{1}{a}\right)}{ln(2)}$ &
\textbf{(4):
$\scriptstyle\frac{13}{2}a+19+\frac{ln\left(\frac{1}{a}\right)}{ln(2)}+2\frac{ln\left(\frac{4a+3}{a}\right)}{ln(2)}$}\\\cline{2-3}
&
$\scriptstyle\frac{9}{2}a-\frac{a_k}{2}+22+\frac{ln\left(\frac{1}{a}\right)}{ln(2)}+\frac{ln\left(\frac{2(a+1)}{a_k}\right)}{ln(2)}$
&
$\scriptstyle\frac{27}{2}a-\frac{a_k}{2}+29+\frac{ln\left(\frac{1}{a}\right)}{ln(2)}+\frac{ln\left(\frac{2(a+1)}{a_k}\right)}{ln(2)}+2\frac{ln\left(\frac{4a+3}{a}\right)}{ln(2)}$(2)
\\\cline{2-3} & & $\scriptstyle
12a-\frac{a_k}{2}+16+\frac{ln\left(\frac{2(a+1)}{a_k}\right)}{ln(2)}+2\frac{ln\left(\frac{4a+3}{a}\right)}{ln(2)}$
(6)\\\cline{2-3} & & \textbf{(5): $\scriptstyle
14a-\frac{a_k}{2}+27+\frac{ln\left(\frac{2(a+1)}{a_k}\right)}{ln(2)}+2\frac{ln\left(\frac{4a+3}{a}\right)}{ln(2)}$}\\\hline
\end{tabular}
\caption{\label{cs1} The upper bounds for the competitive ratio in
the different cases.}
\end{table}

\begin{table}
\centering
\begin{tabular}{|c||c|l|}\hline
$c \le$& $k = 0$ & $k > 0$\\\hline\hline
$B.,$ &
$\scriptstyle 4$ &
\\\cline{2-3} $e \ge a+1,$ & $\scriptstyle
5a-\frac{a_k}{2}+12+\frac{ln\left(\frac{2(a+1)}{a_k}\right)}{ln(2)}$
& \textbf{(6): $\scriptstyle
12a-\frac{a_k}{2}+18+\frac{ln\left(\frac{2(a+1)}{a_k}\right)}{ln(2)}+2\frac{ln\left(\frac{4a+3}{a}\right)}{ln(2)}$}\\\cline{2-3}
$(\alpha)$ & &\textbf{(7): $\scriptstyle
15a-\frac{a_k}{2}+\frac{35}{2}+\frac{ln\left(\frac{4a+3}{a_k}\right)}{ln(2)}+2\frac{ln\left(\frac{4a+3}{a}\right)}{ln(2)}$}\\\cline{2-3}
& $\scriptstyle
-a_k+\frac{47}{2}+2\frac{ln\left(\frac{4a+3}{a_k}\right)}{ln(2)}$ &
$\scriptstyle
21a-a_k+\frac{47}{2}+2\frac{ln\left(\frac{4a+3}{a_k}\right)}{ln(2)}+2\frac{ln\left(\frac{2a+3}{a}\right)}{ln(2)}$
(14) \\\cline{2-3} & $\scriptstyle
8a-a_k+17+2\frac{ln\left(\frac{2(a+1)}{a_k}\right)}{ln(2)}$ &
\textbf{(8): $\scriptstyle
15a-a_k+21+2\frac{ln\left(\frac{4a+3}{a}\right)}{ln(2)}+2\frac{ln\left(\frac{2(a+1)}{a_k}\right)}{ln(2)}$}\\\cline{2-3}
& & $\scriptstyle
12a-a_k+\frac{35}{2}+2\frac{ln\left(\frac{4a+3}{a}\right)}{ln(2)}+\frac{ln\left(\frac{2(a+1)}{a_k}\right)}{ln(2)}$
(10) \\\cline{2-3} & $\scriptstyle
12a-\frac{a_k}{2}+\frac{29}{2}+\frac{ln\left(\frac{4a+3}{a_k}\right)}{ln(2)}$
& \textbf{(9): $\scriptstyle
19a-\frac{a_k}{2}+\frac{43}{2}+\frac{ln\left(\frac{4a+3}{a_k}\right)}{ln(2)}+2\frac{ln\left(\frac{4a+3}{a}\right)}{ln(2)}$}\\\cline{2-3}
&
$\scriptstyle\frac{23}{8}a-\frac{a_k}{4}+\frac{35}{4}+\frac{1}{2}\frac{ln\left(\frac{3(a+1)}{a_k}\right)}{ln(2)}$
& \textbf{(10): $\scriptstyle
\frac{41}{4}a-\frac{a_k}{4}+\frac{35}{4}+2\frac{ln\left(\frac{4a+3}{a}\right)}{ln(2)}+\frac{1}{2}\frac{ln\left(\frac{3(a+1)}{a_k}\right)}{ln(2)}$}\\\cline{2-3}
& $\scriptstyle
7a-\frac{a_k}{4}+\frac{41}{4}+\frac{1}{2}\frac{ln\left(\frac{4a+3}{a_k}\right)}{ln(2)}$
& \textbf{(11): $\scriptstyle
14a-\frac{a_k}{4}+\frac{57}{4}+2\frac{ln\left(\frac{4a+3}{a}\right)}{ln(2)}+\frac{1}{2}\frac{ln\left(\frac{4a+3}{a_k}\right)}{ln(2)}$}\\\cline{2-3}
& $\scriptstyle
-a_k+24+2\frac{ln\left(\frac{4a+3}{a_k}\right)}{ln(2)}$ &
\textbf{(12): $\scriptstyle
21a-a_k+24+2\frac{ln\left(\frac{4a+3}{a}\right)}{ln(2)}+2\frac{ln\left(\frac{4a+3}{a_k}\right)}{ln(2)}$}\\\cline{2-3}
&
$\scriptstyle\frac{7}{2}a-\frac{a_k}{2}+17+\frac{ln\left(\frac{4a+3}{a_k}\right)}{ln(2)}$
& $\scriptstyle
15a-\frac{a_k}{2}+17+2\frac{ln\left(\frac{4a+3}{a}\right)}{ln(2)}+\frac{ln\left(\frac{4a+3}{a_k}\right)}{ln(2)}$
(11) \\\hline
$B.,$ & $\scriptstyle 4$ &
\\\cline{2-3} $e \ge a+1,$ & $\scriptstyle 4$
& \textbf{(13):
$\scriptstyle3a+10+2\frac{ln\left(\frac{a+1}{a}\right)}{ln(2)}$}\\\cline{2-3}
$(\beta)$ & $\scriptstyle
5a+12+\frac{ln\left(\frac{2(a+1)}{a_k}\right)}{ln(2)}$ &
\textbf{(14): $\scriptstyle
8a-\frac{a_k}{2}+16+2\frac{ln\left(\frac{a+1}{a}\right)}{ln(2)}+\frac{ln\left(\frac{2(a+1)}{a_k}\right)}{ln(2)}$}\\\cline{2-3}
&
$\scriptstyle\frac{5}{2}a-\frac{a_k}{4}+6+\frac{1}{2}\frac{ln\left(\frac{2(a+1)}{a_k}\right)}{ln(2)}$
&  $\scriptstyle
\frac{19}{2}a-\frac{a_k}{4}+10+\frac{1}{2}\frac{ln\left(\frac{2(a+1)}{a_k}\right)}{ln(2)}+2\frac{ln\left(\frac{4a+3}{a}\right)}{ln(2)}$
(17) \\\cline{2-3} &
$\scriptstyle\frac{7}{2}a-\frac{a_k}{4}+8+\frac{1}{2}\frac{ln\left(\frac{2(a+1)}{a_k}\right)}{ln(2)}$
&  \textbf{(15): $\scriptstyle
\frac{21}{2}a-\frac{a_k}{4}+10+\frac{1}{2}\frac{ln\left(\frac{2(a+1)}{a_k}\right)}{ln(2)}+2\frac{ln\left(\frac{4a+3}{a}\right)}{ln(2)}$}\\\cline{2-3}
& $\scriptstyle
8a-\frac{a_k}{2}+17+2\frac{ln\left(\frac{2(a+1)}{a_k}\right)}{ln(2)}$
&  \textbf{(16): $\scriptstyle
15a-\frac{a_k}{2}+21+2\frac{ln\left(\frac{2(a+1)}{a_k}\right)}{ln(2)}+2\frac{ln\left(\frac{4a+3}{a}\right)}{ln(2)}$}\\\cline{2-3}
& $\scriptstyle
5a-\frac{a_k}{2}+\frac{25}{2}+\frac{ln\left(\frac{2(a+1)}{a_k}\right)}{ln(2)}$
& \textbf{(17): $\scriptstyle
12a-\frac{a_k}{2}+\frac{33}{2}+\frac{ln\left(\frac{2(a+1)}{a_k}\right)}{ln(2)}+2\frac{ln\left(\frac{4a+3}{a}\right)}{ln(2)}$}\\\hline
$B.,$ & $\scriptstyle
7a-\frac{a_k}{2}+\frac{29}{2}+\frac{ln\left(\frac{2a+3}{a_k}\right)}{ln(2)}$
& \textbf{(18): $\scriptstyle
10a-\frac{a_k}{2}+\frac{43}{2}+2\frac{ln\left(\frac{2a+3}{a_k}\right)}{ln(2)}+2\frac{ln\left(\frac{2a+3}{a}\right)}{ln(2)}$}\\\cline{2-3}
$e < a+1$ & $\scriptstyle
-\frac{a_k}{2}+\frac{35}{2}+\frac{ln\left(\frac{2a+3}{a_k}\right)}{ln(2)}$
& $\scriptstyle
4a-\frac{a_k}{2}+\frac{35}{2}+2\frac{ln\left(\frac{2a+3}{a}\right)}{ln(2)}+\frac{ln\left(\frac{2a+3}{a_k}\right)}{ln(2)}$
(20) \\\cline{2-3} & $\scriptstyle
-a_k+34+2\frac{ln\left(\frac{2a+3}{a_k}\right)}{ln(2)}$ &
\textbf{(19): $\scriptstyle
9a-a_k+34+2\frac{ln\left(\frac{2a+3}{a_k}\right)}{ln(2)}+2\frac{ln\left(\frac{2a+3}{a}\right)}{ln(2)}$}\\\cline{2-3}
& $\scriptstyle
6a-a_k+18+2\frac{ln\left(\frac{a+2}{a_k}\right)}{ln(2)}$ &
$\scriptstyle
9a-a_k+22+2\frac{ln\left(\frac{a+2}{a_k}\right)}{ln(2)}+2\frac{ln\left(\frac{2a+3}{a}\right)}{ln(2)}$
(24)\\\cline{2-3} & $\scriptstyle
4a-a_k+18+2\frac{ln\left(\frac{a+2}{a_k}\right)}{ln(2)}$ &
\textbf{(20): $\scriptstyle
7a-\frac{a_k}{2}+18+\frac{ln\left(\frac{a+2}{a_k}\right)}{ln(2)}+2\frac{ln\left(\frac{2a+3}{a}\right)}{ln(2)}$}\\\hline
\end{tabular}
\caption{\label{cs2} The upper bounds for the competitive ratio in
the different cases.}
\end{table}

\noindent For $a_k = a$, $a \in ]0,1]$, we get the following values:

\[ c \le \left\{
\begin{array}{r@{\quad:\quad}l} 8a+34+4\log\left(2+3/a\right)
& a \in \left]0,0.7004344\right] \\
20a+24+4\log\left(4+3/a\right) & a \in \left]0.7004344,1\right]
\end{array} \right. \]

 \noindent For a given $a$ we get a constant competitive ratio that
depends on $\log (1/a)$. \end{proof}

\noindent Table~\ref{ac} shows the competitive values that strategy
SCANSEARCH achieves for $a_k=a$.
\begin{table}[h]
\centering
\begin{tabular}{|l||r|}\hline
$a$ & upper bound for $c$ \\\hline\hline
$1$ & $55.2294$\\
$0.9$ & $53.2294$\\
$0.8$ & $51.8168$\\
$0.7$ & $50.2083$\\
$0.6$ & $50.0294$\\
$0.5$ & $50.0000$\\
$0.4$ & $50.1917$\\
$0.3$ & $50.7399$\\
$0.2$ & $51.9499$\\
$0.1$ & $54.8000$\\
$0.01$ & $67.0336$\\
$0.001$ & $80.2148$\\
$0.0001$ & $93.4919$\\
$0.00001$ & $106.7785$\\
$0.000001$ & $120.0661$\\\hline
\end{tabular}
\caption{\label{ac}Values for $a$ and the corresponding upper bound
for the competitive ratio.}
\end{table}

\begin{figure}[t!hb]
\centering
\epsfig{file=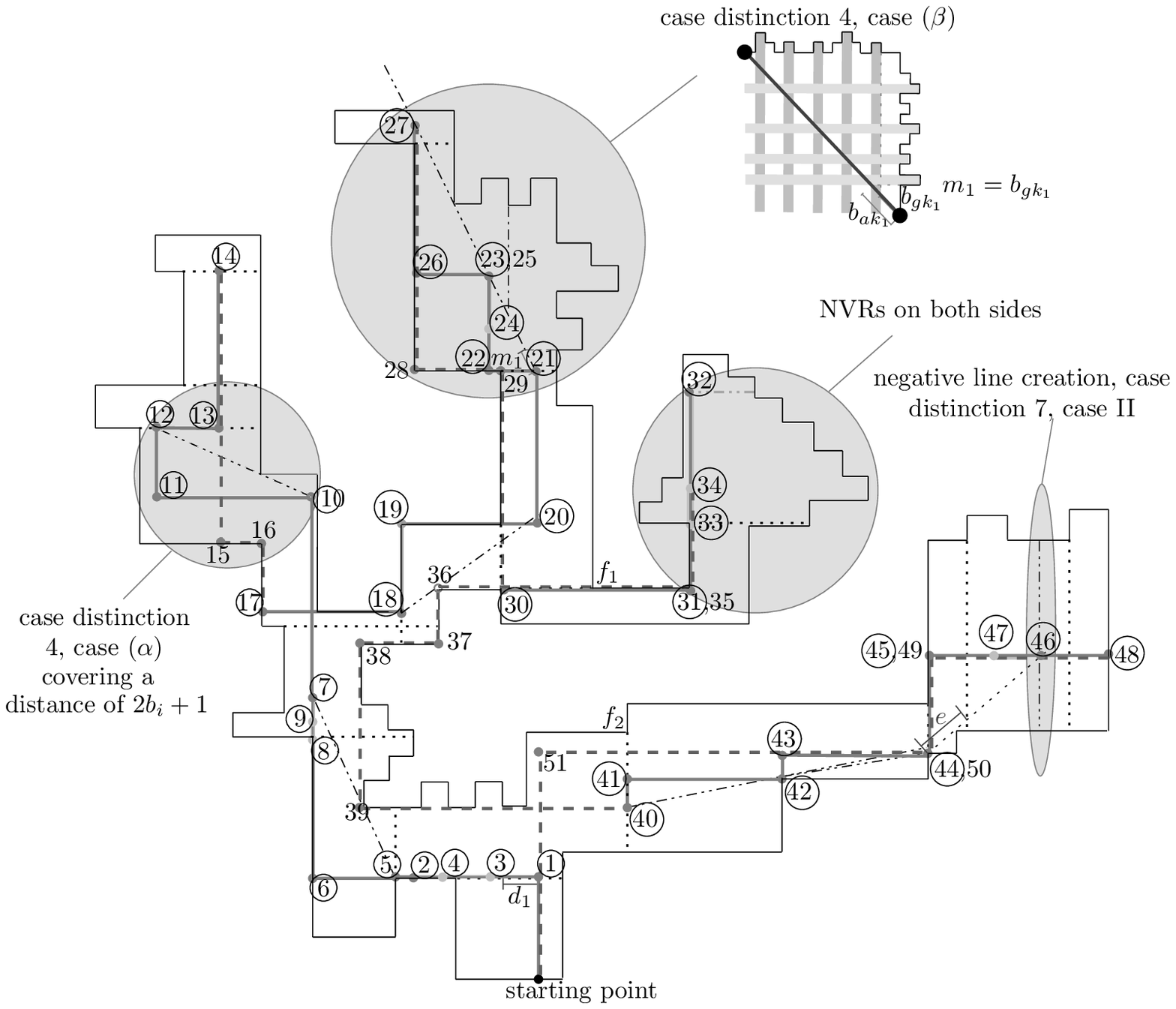, width = .95\textwidth}\\
\caption{\label{Beispiel} An example for strategy SCANSEARCH. The
path of the robot is plotted in gray, we use light gray for binary
searches, and dashed dark gray for parts of the path where it
improves clarity. Extensions are dotted in black, some straight
connections are dotted in light gray. (For clarity, some lines are
slightly offset from their actual position.) Numbered points
correspond to turns in the tour, scan points are circled. }
\end{figure}

See Figure \ref{Beispiel} for an example of our strategy, with $a =
0.5$ $(< 1)$: The starting point is  the black point in the south of
the polygon. The first extension may be reached on a straight,
axis-parallel line, i.e., we are in case ($A.$). As $e \ge 2a+1$
(interval case) and no non-visible regions appear on the
counterclockwise side up to $E$, the robot moves directly to $E$
(point $1$) and takes a scan. Examples of further decisions along
the way are highlighted (scans 9, 20, 30, 45).

\label{subsec:ratio}

\section{Conclusions}
We have considered the online problem of exploring a polygon with a
robot that has discrete vision. In case of a rectilinear
multi-connected polygon we have seen that no strategy may have a
constant competitive ratio; even for orthoconvex polygons, it has
turned out that any bound on the competitive factor must involve the
aspect ratio, even under the assumption that each edge
must be fully visible from some scan point. Finally, we have developed
a competitive strategy for
simple rectilinear polygons under the constraint of full edge visibility.
For this purpose it was important that
we were able to order the extensions along the optimal route of a
robot without continuous vision; this enabled us to compare the cost
of the optimum with the cost of a robot that uses our strategy.

There are several natural open problems. Can we give a competitive
strategy without the assumption that edges are fully visible from
scan points? A prerequisite would be a competitive strategy
for online guard placement.

Another question is whether there exists a competitive strategy in
case of a more general class of regions: simple polygons. In this
context we face the difficulty that we do not know where the
extensions lie. Thus, we are not able to give an a-priori lower
bound on the length of the optimum; this is a serious obstacle to
adapting the algorithm's step length to the one of the optimum;
extending the highly complex method of Hoffmann et al.\ for
continuous vision to the case of discrete vision is an intriguing
and challenging problem.

Finally, it is interesting to consider the offline problem for
various classes of polygons. As stated in the introduction, even the
case of simple rectilinear polygons is NP-hard; developing
reasonable approximation methods and heuristics would be both
interesting in theory as well as useful in practice. In this
context, we have been investigating the situation with the added
constraint of limited visibility.

\small
\bibliographystyle{abbrv}
\bibliography{lit}

\end{document}